\documentclass[review]{elsarticle}
\usepackage{amsmath}
\usepackage{amssymb}
\usepackage{epsfig}
\usepackage{color}
\usepackage{color,soul}
\soulregister\cite7
\soulregister\ref7
\usepackage{textgreek}
\usepackage{float}
\usepackage{subfigure}  
\usepackage{caption}    
\usepackage{algorithm}
\usepackage{algorithmic}
\usepackage{ulem}
\usepackage[pdftex]{hyperref}
\usepackage{graphicx}
\usepackage[titletoc,title]{appendix}
\usepackage{geometry}
\begin{document}
\renewcommand\appendix{\par
  \setcounter{section}{0}
  \setcounter{subsection}{0}
  \setcounter{figure}{0}
  \setcounter{table}{0}
  \renewcommand\thesection{Appendix \Alph{section}}
  \renewcommand\thefigure{\Alph{section}\arabic{figure}}
  \renewcommand\thetable{\Alph{section}\arabic{table}}
}

\begin{frontmatter}
\title{A Consistent Multi-Resolution Smoothed Particle Hydrodynamics Method}

\author[BIT]{Wei Hu}
\author[WISC]{Wenxiao Pan\corref{cor}}
\ead{wpan9@wisc.edu}
\author[WISC]{Milad Rakhsha}
\author[BIT]{Qiang Tian}
\author[BIT]{Haiyan Hu}
\author[WISC]{Dan Negrut}

\cortext[cor]{Corresponding author}
\address[BIT]{School of Aerospace Engineering, Beijing Institute of Technology, Beijing, 100081, China}
\address[WISC]{Department of Mechanical Engineering, University of Wisconsin-Madison, Madison, WI 53706, USA}

\begin{abstract}
We seek to accelerate and increase the size of simulations for fluid-structure interactions (FSI) by using multiple resolutions in the spatial discretization of the equations governing the time evolution of systems displaying two-way fluid-solid coupling. To this end, we propose a multi-resolution smoothed particle hydrodynamics (SPH) approach in which subdomains of different resolutions are directly coupled without any overlap region. The second-order consistent discretization of spatial differential operators is employed to ensure the accuracy of the proposed method. As SPH particles advect with the flow, a dynamic SPH particle refinement/coarsening is employed via splitting/merging to maintain a predefined multi-resolution configuration. Particle regularity is enforced via a particle-shifting technique to ensure accuracy and stability of the Lagrangian particle-based method embraced. The convergence, accuracy, and efficiency attributes of the new method are assessed  by simulating four different flows. In this process, the numerical results are compared to the analytical, finite element, and consistent SPH single-resolution solutions. We anticipate that the proposed multi-resolution method will enlarge the class of SPH-tractable FSI applications.
\end{abstract}

\begin{keyword}
smoothed particle hydrodynamics \sep multi-resolution \sep refinement and coarsening  \sep fluid-structure interactions
\end{keyword}

\end{frontmatter}

\section{Introduction}\label{sec:Introduction}

SPH is a Lagrangian method that has been used to solve partial differential equations (PDEs) associated with mass, momentum, and energy conservation laws \cite{Monaghan_SPH_2005}. In SPH, these PDEs are spatially discretized employing a set of particles that possess material properties and interact with each other in a fashion controlled by a smoothing, or kernel, function with compact support. These particles move according to inter-particle interactions and external forces. Due to its Lagrangian nature, SPH enables efficient modeling of multiphase flows with moving interfaces \cite{Adami2010surface,Tofighi2013numerical,Tart2016JCP}, complex fluids  \cite{PanJG2012,PanJCP2013,SPHPolymer2016,SPHNoncolloid2016}, materials subject to large deformation \cite{Gray2001elastic,Pan2013material,Hu2016dynamic}, fluid-structure interactions (FSI) problems  \cite{Hu2014dynamic,Schorgenhumer2013interaction,Pazouki2014FSI}, and various transport phenomena \cite{Pan_LLNS_2014,Tart2015CompGeo,PanBMC2015,SPHProton2015}. 

In most SPH simulations, the computational domain is populated with particles of uniform spacing, which leads to a single--resolution scenario. This is suitable when there are relatively small variations in the velocity/pressure gradient over the entire domain of analysis. Otherwise, it is desirable to discretize the domain with SPH particles of multiple resolutions. For instance, for a colloids-in-suspension FSI problem, the near field, i.e., next to each colloid, would call for high resolution; the far field would rely on a coarser resolution. Due to the Lagrangian nature of SPH, a locally refined region is expected to move along with the immersed body leading to an adaptive particle refinement (APR) process. In this regard, APR for SPH serves the same purpose as the adaptive mesh refinement does in the finite element and finite volume methods for FSI. The expectation is that APR will maintain the accuracy of the SPH solution while improving efficiency and increasing the size of the problem solved.

In this context, several multi-resolution SPH methods have been proposed, e.g., \cite{Lastiwka2009,Oger2006wedge,Feldman2007refinement,Lopez2013refinement,Vacondio2013variable,Barcarolo2014refinement,Bian2015DDSPH}. In these efforts, SPH particles were distributed according to a pre-defined multi-resolution configuration. What set these approaches apart was the coupling between regions of different resolutions and thereby the accuracy and continuity of the numerical solution at and near the interface. In \cite{Omidvar2012variable2D}, the same large smoothing length was used for both the low- and high-resolution regions. This enhanced the accuracy attribute of the solution in some degree but led to a large number of neighbors for the particles in the high-resolution region. Thus, performance was hurt, which defeated APR's purpose. In \cite{Vacondio2013variable,Vacondio2016variable,Barcarolo2014refinement}, each region had its own smoothing length, which reduced the computational cost substantially. However, using an inconsistent SPH discretization, the numerical solution near the interface lacked accuracy, an artifact that became worse as the ratio of the two resolutions increased. This point will be demonstrated in \S\ref{subsec:Poi_Flow}. To tackle this issue, an approach promoting different-resolution subdomains connected via an overlap region was described in \cite{Bian2015DDSPH}. Therein, the overlap region was employed to exchange state information and fluxes. At each time step, the solutions in the two subdomains were brought in sync via the overlap region by an iterative process. The authors investigated the sensitivity of numerical accuracy with respect to the thickness of the overlap region; a minimum thickness of the overlap region was determined numerically. On the upside, the approach can employ a different time step for each of the subdomains, hence reducing the computational effort in the low-resolution subdomain. On the downside, bringing the solutions of the two subdomains in agreement via an iterative algorithm remains costly and ad-hoc. Finally, if the two subdomains are moving dynamically, continually establishing an adaptive overlap region poses further difficulties.
 
Therefore, we propose a different multi-resolution SPH method in this work. In that, each region has its own smoothing length, and particles of different resolutions directly interact without any overlap region. The second-order consistent discretization is employed to discretize the governing equations via local correction matrices computed at each particle \cite{Fatehi2011error,Traska2015second-order,Pan2017Modeling}. It thereby ensures the accuracy and continuity of the solution at the interface. Due to its higher-order accuracy and convergence, our multi-resolution SPH needs fewer number of particles to achieve the same accuracy and thereby further saves computational cost. In addition, we employ a dynamic refinement/coarsening of SPH particles to maintain the predefined multi-resolution configuration as SPH particles advect with the flow. In this respect, we follow in the steps of several other dynamic refinement/coarsening approaches \cite{Meglicki19933d,Kitsionas2002splitting,Kitsionas2007clump,Feldman2007refinement,Vacondio2012split,Lopez2013refinement,Vacondio2013variable,Barcarolo2014refinement,Vacondio2016variable}. The basic underlying idea is that of splitting a ``big'' particle into several ``small'' ones according to a certain geometric arrangement centered at the location of the big particle and, conversely, merging nearest small particles into a big one at the center-of-mass of those small particles. To a large extent, what differentiates different approaches is how the splitting and merging occur, for example, upon splitting where the small particles are located and how their masses and kernel lengths are assigned.

Against this backdrop, there are several aspects in which the proposed method is different from the current dynamic refinement/coarsening techniques. First, when a big particle moves out of a low-resolution region, we split it into the same count of $n$ small particles according to the ratio of two resolutions. After splitting, the big particle vanishes. In \cite{Barcarolo2014refinement} the big particle is persistent albeit virtual, and its motion through the high-resolution region is calculated from interpolated information provided by the small particles. When the big particle crosses back into the low-resolution region, it sheds its virtual status to become a SPH particle whose motion is dictated by the governing equations. By avoiding virtual particles, our approach requires much simpler code implementation, uses less memory, and eliminates the overhead of keeping tabs on virtual particles. Second, in \cite{Feldman2007refinement}, a big particle was split into small particles with non-uniform masses but the same inter-particle spacing and smoothing length. Therein, the mass of each small particle was determined by minimizing the splitting-induced error in the density field, which only partially alleviated the discontinuity of the solution at the interface between regions of different resolutions. Alternatively, an approach for splitting a big particle into small particles with the same mass was proposed in \cite{Lopez2013refinement}. Therein, the inter-particle spacing and smoothing length were determined by minimizing the splitting-induced error in the density field. In this vein, our approach falls in-between: all small particles have equal mass, spacing $\Delta x_{H}$, and smoothing length $h_{H}$ in the splitting process. Moreover, we maintain the accuracy and continuity of the solution at the interface by means of: ($i$) the consistent SPH discretization and ($ii$) interpolation-based splitting/merging process. In ($ii$), we use a second-order interpolation scheme to endow each small particle with its own velocity, with a similar interpolation-based approach used for merging. In this regard, our approach is more accurate than the averaging-based technique presented in \cite{Vacondio2013variable,Vacondio2016variable}. Finally, compared to these latter techniques, we always split or merge the same number $n$ of particles, which leads to the presence of only two families of SPH particles at the low-resolution/high-resolution interface. 

Furthermore, in Lagrangian particle-based methods, particle regularity is critical to ensure numerical stability and accuracy of the simulation. However, highly distorted particle distribution may appear as SPH particles advect with flow. Thus, in this method, we employ a particle shifting technique to enforce a uniform particle distribution without sacrificing efficiency and Lagrangian nature. In that, a particle is slightly shifted away from its streamline according to a computed shifting vector, and after shifting hydrodynamic variables are corrected according to its new positions using the second-order interpolation. The multi-resolution configuration of particles is particularly considered in determining the shifting vector. Ultimately, the no-slip boundary condition for the fluid velocity is imposed through ghost solid particles with velocities linearly extrapolated from the fluid velocity \cite{TakedaSPHBC1994,Morris1997modeling,Holmes2011Smooth}, which ensures the second-order accuracy at the boundaries \cite{Macia2011SPHBCs}.

This contribution is organized as follows. The next section provides a detailed account of the proposed new multi-resolution SPH method. To that end, it touches on several topics: standard SPH discretization, consistent SPH discretization, dynamic splitting and merging of SPH particles, particle-shifting technique, enforcement of the no-slip boundary condition, and time integration. In Section \ref{sec:Simu_results}, we demonstrate the accuracy, convergence, and efficiency of the proposed method through numerical tests. We model four different flows -- the transient Poiseuille flow, flow around a periodic array of cylinders, flow around a rotating cylinder near a moving wall, and a translating and rotating ellipsoid moving in fluid. The numerical results obtained are compared with analytical, finite element, or consistent SPH single-resolution solutions. We provide concluding remarks and directions of future work in Section \ref{sec:conclusion}.


\section{Multi-resolution SPH method}\label{sec:Num_Meth}

\subsection{Lagrangian hydrodynamic equations}

We consider the fluid flow taking place in the space $\Omega_f$ and the solid (wall or immersed body/bodies) occupying the space $\Omega_s$. The boundary $\Gamma$ separates $\Omega_f$ and $\Omega_s$, i.e., $\Gamma = \Omega_f \cap \Omega_s$. The time evolution of the fluid is governed by the continuity and Navier-Stokes (NS) equations
\begin{equation}
\begin{cases}
       \frac{d\rho}{dt}=-\rho \: \nabla \cdot \textbf{v} \\
       \frac{d\textbf{v}}{dt}=-\frac{1}{\rho} \nabla p + \nu \nabla^{2} \textbf{v} + \textbf{f}_{body} \; ,
\end{cases} \\
\label{equ:NS_momentum}
\end{equation}
which hold for $\textbf{x}\in \Omega_f$. Here, $p$ is the pressure; $\textbf{f}_{body}$ is the body force per unit mass acting on the fluid; $\nu = \frac{\mu}{\rho}$ is the kinematic viscosity of the fluid; $\rho$ and $\mu$ are fluid density and viscosity, respectively. In the weakly compressible SPH, the above equations are closed by an equation of state (EOS) for the pressure $p$ as:
\begin{equation}\label{equ:state-equ}
p= c^2 \rho,
\end{equation}
where $c$ is the speed of sound. This EOS guarantees nonzero background pressure and hence no negative pressure in the fluid field \cite{Balsara1995StaSPH,Swegle1995StaSPH,Morris1996StaSPH}.
At the fluid-solid boundary, the no-slip boundary condition for the NS equation is imposed as $\textbf{v} = \mathbf{v}_\Gamma$ for $\forall~\textbf{x}\in \Gamma$, where $\mathbf{v}_\Gamma$ is the velocity of the boundary $\Gamma$.

\subsection{Spatial discretization}\label{subsec:SPH Cons_Dis}  
Herein, we rely on SPH for the spatial discretization of the equations of motion.
%
The value of a function $f$ at $\textbf{x}_i$, the position of particle $i$, is then approximated as \cite{Monaghan_SPH_2005}:
\begin{equation}\label{equ:sph_fun}
f_i = \sum\limits_{j}f_j W_{ij}V_j.
\end{equation}
Here, $V$ is the volume of a SPH particle and defined as:
\begin{equation}\label{equ:sph_vol}
V_i = (\sum\limits_{j}W_{ij})^{-1} \; ,
\end{equation}
where $W_{ij} = W(\textbf{r}_{ij})$ is the kernel function. We adopt a quintic spline function $W$ that has continuous and smooth first and second derivatives, and which is expressed as
\begin{equation}\label{equ:sph_ker}
W_{ij} = \alpha_d
\begin{cases}
\ (3-R)^5 - 6(2-R)^5 + 15(1-R)^5 ~~~0 \leqq R < 1 \\
\ (3-R)^5 - 6(2-R)^5 ~~~~~~~~~~~~~~~~~~~~~1 \leqq R < 2 \\
\ (3-R)^5 ~~~~~~~~~~~~~~~~~~~~~~~~~~~~~~~~~~~~~2 \leqq R < 3 \\
0 ~~~~~~~~~~~~~~~~~~~~~~~~~~~~~~~~~~~~~~~~~~~~~~~~~~~~~ R \geqq 3. \\
\end{cases} \\
\end{equation}
Here, $R=\frac{r_{ij}}{h}$; $h$ is the kernel length; $r_{ij}=|\textbf{r}_{ij}|$; $\textbf{r}_{ij} = \textbf{x}_i - \textbf{x}_j$; and, for future reference, $\textbf{e}_{ij}=\textbf{r}_{ij}/r_{ij}$. The constant $\alpha_d$ assumes the value $120/h$, $7/478\pi h^2$ and $3/359\pi h^3$ for 1-, 2- and 3D problems, respectively. Only neighbor particles of particle $i$ in the set ${\cal{N}}_{h,i} = \left\{ \textbf{x}_j~ \text{s.t.}~ |\textbf{x}_j - \textbf{x}_i| < 3h \right\}$ contribute to the summation in the interpolation scheme of Eq. \eqref{equ:sph_fun}, i.e., $j \in {\cal{N}}_{h,i}$. According to this equation, the density at $\textbf{x}_i$ can be expressed as
\begin{equation}\label{equ:con_sum}
\rho_i = \sum\limits_{j} m_j W_{ij},
\end{equation}
with $m_j$ denoting the mass of particle $j$. 

\begin{figure}[H]
\centering
\includegraphics[scale=0.45]{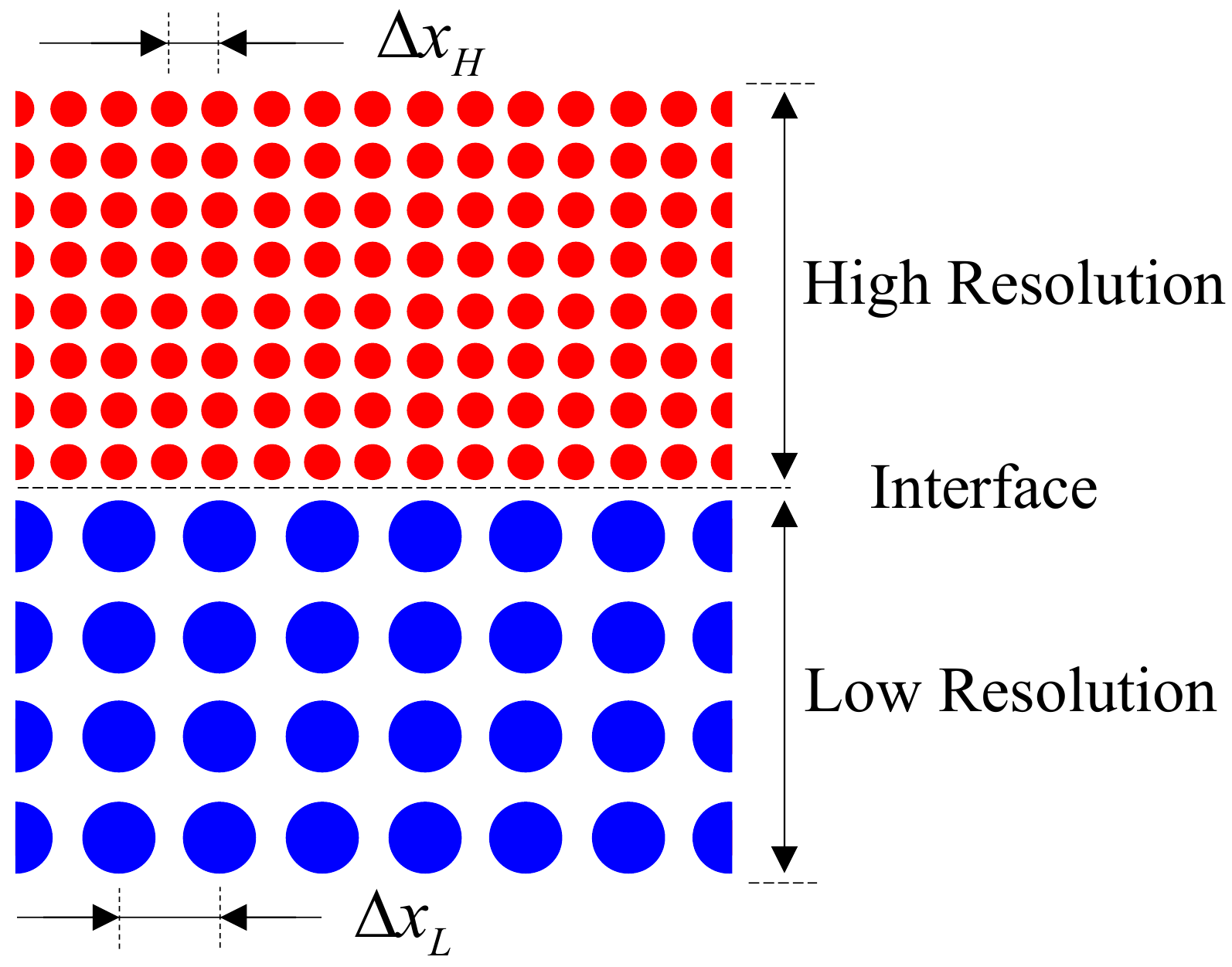}
\caption{Initial particle configuration in the multi-resolution SPH method.}
\label{fig:Mul-Reso-Conf}
\end{figure}

In the proposed multi-resolution SPH method, the computational domain is divided into two regions: a high resolution one, which hosts SPH particles with a small $h_H$, and a low resolution one in which the SPH particles have a large $h_L$. As shown in Figure \ref{fig:Mul-Reso-Conf}, all SPH particles are initially placed on a regular lattice with the spacing of $\Delta x_{H}$ in the high-resolution region and the spacing of $\Delta x_{L}$ in the low-resolution region. Different ratios of $\Upsilon_{LH} \equiv \Delta x_L/\Delta x_H$ are allowed in this method. Here, $h_H = 1.25 \Delta x_{H}$, and $h_L = 1.25 \Delta x_{L}$. 

In the standard SPH discretization, the gradient and Laplacian of the function $f_i$ are discretized as
\begin{equation}\label{equ:dis_inconsistent}
\nabla f_{i}=\sum\limits_j (f_{j} -f_{i} )\nabla_i W_{ij}V_{j}  ~~~\mathrm{and}~~~\nabla^{2} f_{i}=2\sum\limits_j (\frac{f_{i}-f_{j}}{r_{ij}}) (\mathbf{e}_{ij} \cdot \nabla_i W_{ij}) V_{j},
\end{equation}
two formulas used in the NS equation for the pressure and viscous terms, respectively. Previous studies have demonstrated that this discretization cannot guarantee the exact gradient for linear functions and Laplacian for parabolic functions and hence introduces inconsistent convergence in the numerical accuracy, i.e., the second-order accuracy of SPH can only be recovered by taking sufficiently large compact support of neighbors \cite{Quinlan2006truncation,Basa2009robustness,Fatehi2011error}. We also noticed that for multi-resolution SPH, this discretization leads to inaccurate numerical solution near the interface. In fact, as shown in Section \ref{subsec:Poi_Flow}, the quality of the solution at the interface will deteriorate as the ratio of two resolutions increases. We alleviate this numerical artifact by using a consistent discretization in which the gradient and Laplacian are evaluated as
\begin{equation}\label{equ:gra_ope}
\nabla f_{i}=\sum\limits_j (f_{j}-f_{i})\textbf{G}_{i}\nabla_i W_{ij}V_{j},
\end{equation}
\begin{equation}\label{equ:lap_ope}
\nabla^{2} f_{i}=2\sum\limits_j [ \textbf{L}_{i}:(\textbf{e}_{ij} \otimes \nabla_i W_{ij})] (\frac{f_{i}-f_{j}}{r_{ij}} - \textbf{e}_{ij}\cdot\nabla f_{i}) V_{j},
\end{equation}
where ``$\otimes$" represents the dyadic product of the two vectors, and ``$:$" represents the double dot product of two matrices. The correction matrices $\textbf{G}_i$ and $\textbf{L}_i$ guarantee the exact gradient for linear functions and Laplacian for parabolic functions regardless of the ratio of $h/\Delta x$ \cite{Fatehi2011error,Traska2015second-order,Pan2017Modeling}. Hence, the consistent discretization of the NS equation is given as:
\begin{equation}\label{equ:NS_discretization}
\frac{d\mathbf{v}_i}{dt} = \mathbf{f}_i = -\sum\limits_j \frac{1}{\rho_j} (p_{j}-p_{i})\textbf{G}_{i}\nabla_i W_{ij}V_{j} + 2\sum\limits_j \nu_j [ \textbf{L}_{i}:(\textbf{e}_{ij} \otimes \nabla_i W_{ij})]\\
                 (\frac{\mathbf{v}_{i}-\mathbf{v}_{j}}{r_{ij}} - \textbf{e}_{ij}\cdot\nabla \mathbf{v}_{i}) V_{j} + \mathbf{f}_{body, i}.
\end{equation}
Here, both $\textbf{G}_i$ and $\textbf{L}_i$ are symmetric $n\times n$ matrices in $\mathbb{R}^{n}$. Specifically, the $mn$ component of the inverse matrix of $\textbf{G}_i$ is:
\begin{equation}\label{equ:gt_inv}
(\textbf{G}_{i}^{-1})^{mn}=-\sum\limits_j r_{ij}^{m}\nabla_{i,n}W_{ij}V_{j}.
\end{equation}
To define $\textbf{L}_{i}$, we adopt the Einstein summation convention and seek a solution of the following \cite{Traska2015second-order}:
\begin{equation}\label{equ:delta_mn}
-\delta^{mn}=\sum\limits_j(A_{i}^{kmn}e_{ij}^{k}+r_{ij}^{m}e_{ij}^{n})(L_{i}^{op}e_{ij}^{o}\nabla_{i,p}W_{ij}V_{j}),
\end{equation}
where the third-order tensor $A_{i}^{kmn}$ is defined as:
\begin{equation}\label{equ:ai_kmn}
A_{i}^{kmn}=G_{i}^{kq}\sum\limits_j r_{ij}^{m} r_{ij}^{n}\nabla_{i,q}W_{ij}V_{j}.
\end{equation}
Here, $r_{ij}^{k}$,~$e_{ij}^{k}$,~$x_{i}^{k}$, and $\nabla_{i,k}W_{ij}$ denote the $k$th component of vectors $\textbf{r}_{ij}$,~$\textbf{e}_{ij}$, $\textbf{x}_{i}$, and $\nabla_{i}W_{ij}$, respectively. The details about how the components of $\textbf{G}_i$ and $\textbf{L}_i$ are computed in 2D are presented in Appendix. 

Finally, we employ a renormalization step for calculating the density of SPH particles near the interface of two regions of different resolutions:
\begin{equation}\label{equ:density_renor}
\rho_i = \frac{\sum\limits_{j} m_j W_{ij}}{\sum\limits_{j} V_{j} W_{ij}} \; .
\end{equation}

\subsection{Splitting and merging of SPH particles}\label{subsec:Split_Merge}  

\begin{figure}[H]
\centering
\includegraphics[scale=0.45]{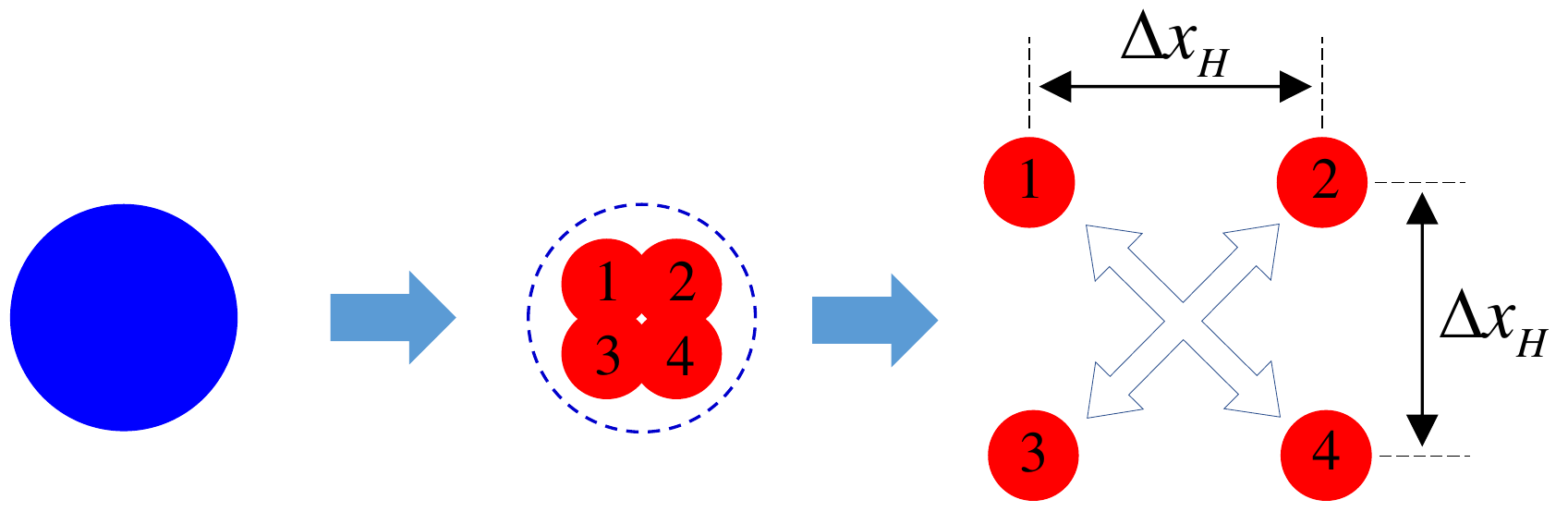}
\caption{Refinement in 2D with $\Upsilon_{LH}=2$; i.e., splitting a big particle into $n=4$ small particles.}
\label{fig:Split-Process}
\end{figure}

Due to the Lagrangian nature of SPH, particles advect with the flow. Thus, as the system evolves, particles from the high-resolution region might leave it to enter a low-resolution region, and vice-versa. This process is anchored by rules facilitating the dynamic splitting and merging of SPH particles. Specifically, when a big particle moves out of its original region, it is split into $n$ small particles following the procedure described in Algorithm 1.
\begin{algorithm}[H] 
\caption{Procedure for refinement via splitting of an SPH particle (see also Fig.~\ref{fig:Split-Process}).}
\begin{algorithmic}[1]
\STATE At its position, replace a big particle with $n$ small ones, each having the velocity and density of the big particle. 
\STATE Move the small particles apart such that they have an equal spacing of $\Delta x_{H}$. 
\STATE  \label{corr-step} Correct the velocity and density of each small particle via Eq.~(\ref{equ:correct_var}) -- see below.
\STATE Renormalize the densities of small particles via Eq.~\eqref{equ:density_renor}. 
\STATE Update the pressures of small particles via Eq. \eqref{equ:state-equ}.
\end{algorithmic}
\end{algorithm}
Note that in the third step above, based on its new position, each new particle receives a small velocity and density correction via a second-order interpolation that draws on information associated with the original ``big'' particle:
\begin{equation}\label{equ:correct_var}
\psi_{i}^{new} = \psi_{i} + \nabla \psi_{i}\cdot \delta \textbf{r}_{i} \; ,
\end{equation}
where the field variable $\psi_{i}$ represents $\textbf{v}_{i}$ or $\rho_{i}$,  and $\nabla \psi_{i}$ is calculated according to Eq.~\eqref{equ:gra_ope}. The second-order accuracy of this interpolation scheme matches that of the consistent SPH discretization (Eqs. \eqref{equ:gra_ope}--\eqref{equ:density_renor}) described above. Figure \ref{fig:Split-Process} further illustrates this refinement procedure in 2D with $\Upsilon_{LH}=2$; i.e., splitting a big particle into $n=4$ small particles. The salient attribute of this approach is linear momentum preservation.

Conversely, when small particles move out of a high-resolution region, the $n$ nearest ones are grouped and merged into a big particle following the procedure described in Algorithm 2. 
\begin{algorithm}[H] 
\caption{Procedure for coarsening via merging of SPH particles (see also Fig.~\ref{fig:Merge-Process}).}
\begin{algorithmic}[1]
\STATE Group $n$ nearest small particles and merge them into a big particle positioned at center of mass of $n$ small particles. 
\STATE Extrapolate densities and velocities of small particles at the location of the big particle via Eq. \eqref{equ:correct_var}.
\STATE Average the $n$ extrapolated densities and velocities to endow the big particle with corresponding values.
\STATE Renormalize the density of the big particle via Eq.~\eqref{equ:density_renor}. 
\STATE Update the pressure of the big particle via Eq. \eqref{equ:state-equ}.
\end{algorithmic}
\end{algorithm}
\begin{figure}[H]
\centering
\includegraphics[scale=0.45]{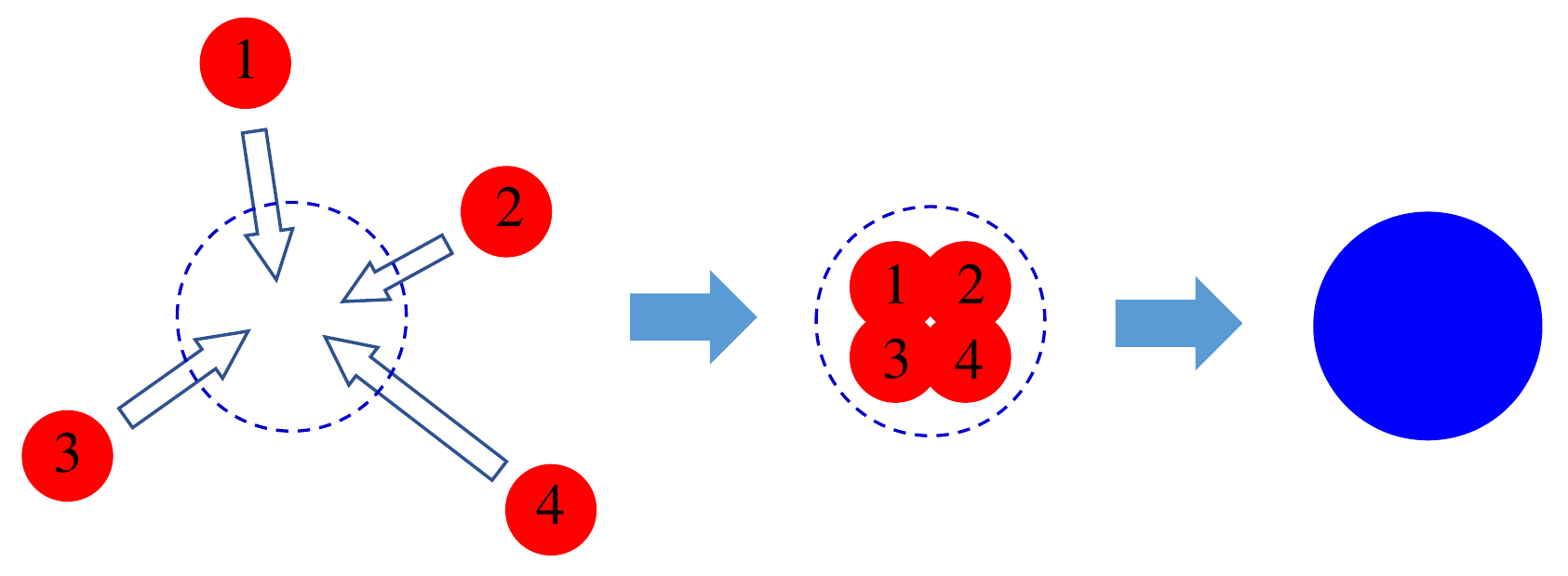}
\caption{Coarsening process in 2D with $\Upsilon_{LH}=2$; i.e., merging $n=4$ small SPH particles into a big one.}
\label{fig:Merge-Process}
\end{figure}
%

If the high/low resolution regions are predefined, the proposed splitting and merging techniques can maintain their geometry over time. In this regard, the techniques proposed are different from the approach in \cite{Vacondio2013variable} in one key aspect. After a big particle is split into small particles, in \cite{Vacondio2013variable} the velocity of each small particle is simply set to that of the big particle without any correction step. Merging occurs gradually with two nearest small particles merged at each time step. Consequently, particles of various sizes co-exist in the computational domain, making it difficult to control the numerical accuracy and to implement the periodic boundary condition.

\subsection{Particle regularity}\label{subsec:Par_Shift}  

SPH particles' advection can lead to scenarios characterized by high particle disorder and/or regions with high particle depletion/plenitude. We provision against such scenarios, which can undermine the accuracy and stability attributes of the numerical solution, by slightly shifting particles away from streamlines to enforce a uniform particle distribution. The particle shifting technique in incompressible SPH was promoted in \cite{Xu2009Accuracy} and subsequently found to be effective in sustaining particle regularity and numerical stability in \cite{Traska2015second-order,Xu2009Accuracy,Pan2017Modeling}. This shifting technique was also applied in a weakly compressible SPH (WCSPH) formulation \cite{Shadloo2012Robust}, and subsequently in a multi-resolution WCSPH method \cite{Vacondio2013variable}. We follow the latter approach, in which the shifting vector $\delta \textbf{r}_{i}$ is determined as
\begin{equation}
\delta \textbf{r}_{i} = \frac{\beta r_0^2 v_\mathrm{max} \Delta t}{{\bar m}_i}\sum\limits_j  m_j\frac{\textbf{r}_{ij}}{r_{ij}^3} \; ,
\label{equ:shifting_vector}
\end{equation}
where $r_0=\frac{1}{N_i}\sum\limits_j  {r_{ij}}$; $N_i = |{\cal{N}}_{h,i}|$; ${\bar m}_i = \sum\limits_j  m_j$; $\beta$ is an adjustable dimensionless parameter and set as $\beta=0.5$ in this work; $v_\mathrm{max}$ is the maximum velocity of the system; $\Delta t$ is the time step of simulation. We introduce the particle's mass in determining the shifting vector in Eq. \eqref{equ:shifting_vector} to maintain the anisotropic configuration of particles near the interface of multi-resolution regions. With this, at the end of each time step, the position of particle $i$ is shifted by
\begin{equation}
\textbf{x}_{i}^{new} = \textbf{x}_{i} + \delta \textbf{r}_{i} \; .
\end{equation}
Accordingly, the $\rho_{i}$ and $\textbf{v}_{i}$ field variables are corrected to the values at the shifted new position via Eq.~\eqref{equ:correct_var}: $\rho_{i} \rightarrow \rho_{i}^{new}$, and ${\bf v}_{i} \rightarrow {\bf v}_{i}^{new}$. For consistency, the pressure $p_{i}^{new}$ is also corrected using $\rho_{i}^{new}$ via the EOS (Eq.~\eqref{equ:state-equ}). 

\subsection{Imposition of no-slip boundary conditions}\label{subsec:Bou_Cons}  

\begin{figure}[H]
    \centering
    \includegraphics[scale=0.40]{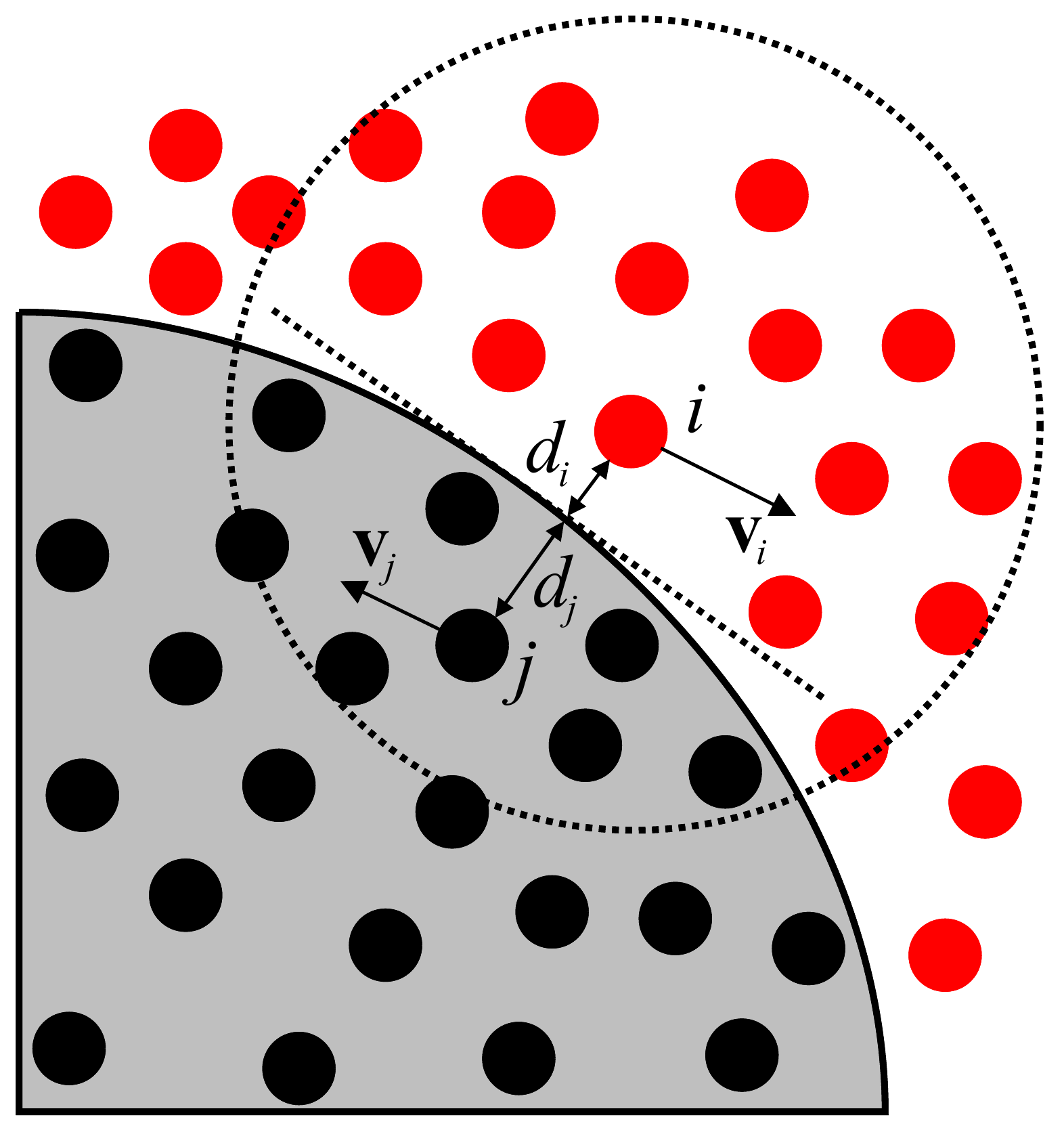}
    \caption{No-slip boundary condition.}
    \label{fig:No-slip}
\end{figure}

We enforce the no-slip boundary condition for the fluid velocity via several layers of ghost (fixed) particles (black) in the solid, as illustrated in Figure \ref{fig:No-slip}. As such, the SPH approximations of velocity and its spatial derivatives for the fluid particles (red) near the fluid-solid boundary have full support of the kernel contained in the domain ($\Omega_f \cup \Omega_s$). As in previous studies \cite{TakedaSPHBC1994,Morris1997modeling}, we perform a linear extrapolation of velocities to the ghost particles. If $i$ is a fluid particle and $j$ a solid ghost particle, the velocity of the latter is computed as
\begin{equation*}
\textbf{v}_j = \frac{d_{j}}{d_{i}}(\textbf{v}_B - \textbf{v}_i) + \textbf{v}_{B} \; ,
\end{equation*}
where $d_{i}$ and $d_{j}$ denote the closest perpendicular distances to the boundary for the fluid and solid particles, respectively; and, $\textbf{v}_{B}$ is the velocity at point $j$ of the solid boundary. As discussed in \cite{Macia2011SPHBCs}, the linear extrapolation introduces a local $\mathcal{O} (h^2)$ error, which matches the second-order accuracy of the consistent SPH. 
There are two caveats. First, to ensure numerical stability, following a recommendation made in \cite{Morris1997modeling}, we cap the relative velocity as 
\begin{equation*}
\textbf{v}_{ij} = \textbf{v}_i -\textbf{v}_j = \textbf{v}_{ij} = \min \{\lambda_{max},1+\frac{d_{j}}{d_{i}}\} \times (\textbf{v}_i - \textbf{v}_B) \; ,
\end{equation*}
with $\lambda_{max}=1.5$. Second, computing $d_i$ and $d_j$ for arbitrary complex geometries can be challenging. Thus, we approximate these distances as in \cite{Holmes2011Smooth}. To that end, an indicator function is used to differentiate fluid and solid particles as:
\begin{equation*}
\chi_i = \frac{\sum\limits_{k\in \mathrm{fluid} }W_{ik}}{\sum\limits_{k}W_{ik}} \; , \qquad \qquad
\chi_j = \frac{\sum\limits_{k\in \mathrm{solid} }W_{jk}}{\sum\limits_{k}W_{jk}}.
\end{equation*}
The distances to the boundary are then approximated as
\begin{equation*}
d_i = \omega h_{i} (2\chi_i-1) \; , \qquad \qquad
d_j = \omega h_{j} (2\chi_j-1) \; ,
\end{equation*}
where for the quintic spline function used in this work the kernel compact support is $\omega=3$. This approach allows the specification of an arbitrarily complex geometry by simply placing a lattice of particles over a domain $\Omega$ and labeling particles as either fluid or solid.

\subsection{Time integration scheme}\label{subsec:Time_Integ}  
Time integration is performed using a second-order, explicit predictor-corrector scheme \cite{Monaghan1989problem,Monaghan1994simulating}. If $t$ and $t+\Delta t$ denote the current and next time steps, respectively, a particle's velocity $\bar{\textbf{v}}_{i}$ and position $\bar{\textbf{x}}_{i}$ at the intermediate time step ($t+\frac{1}{2}\Delta t$) are first predicted as:
\begin{equation*}
\begin{cases}
\ \bar{\textbf{v}}_{i}^{t+\frac{1}{2}\Delta t} = \textbf{v}_{i}^{t} + \frac{1}{2}\Delta t \textbf{f}_{i}^{t}, \\
\ \bar{\textbf{x}}_{i}^{t+\frac{1}{2}\Delta t} = \textbf{x}_{i}^{t} + \frac{1}{2}\Delta t \textbf{v}_{i}^{t} \;, \\
\end{cases} \\
\end{equation*}
where $\textbf{f}_{i}$ is the total force per unit mass exerted on particle $i$. The corresponding density $\bar{\rho}_{i}^{t+\frac{1}{2}\Delta t}$ and pressure $\bar{p}_{i}^{t+\frac{1}{2}\Delta t}$ are obtained via Eqs.~\eqref{equ:con_sum} and \eqref{equ:state-equ}, respectively. Using this intermediate-step information, $\textbf{f}_{i}^{t+\frac{1}{2}\Delta t}$ is evaluated and subsequently used for correction purposes:
\begin{equation*}
\begin{cases}
\ \textbf{v}_{i}^{t+\frac{1}{2}\Delta t} = \textbf{v}_{i}^{t} + \frac{1}{2}\Delta t \textbf{f}_{i}^{t+\frac{1}{2}\Delta t}, \\
\ \textbf{x}_{i}^{t+\frac{1}{2}\Delta t} = \textbf{x}_{i}^{t} + \frac{1}{2}\Delta t \textbf{v}_{i}^{t+\frac{1}{2}\Delta t}. \\
\end{cases} \\
\end{equation*}
Finally, the particle's velocity and position at $t+\Delta t$ are computed as
\begin{equation*}
\begin{cases}
\ \textbf{v}_{i}^{t+\Delta t} = 2\textbf{v}_{i}^{t+\frac{1}{2}\Delta t} - \textbf{v}_{i}^{t}, \\
\ \textbf{x}_{i}^{t+\Delta t} = 2\textbf{x}_{i}^{t+\frac{1}{2}\Delta t} - \textbf{x}_{i}^{t} \; , \\
\end{cases} \\
\end{equation*}
while the density $\rho_{i}^{t+\Delta t}$ and pressure $p_{i}^{t+\Delta t}$ are obtained via Eqs.~\eqref{equ:con_sum} and \eqref{equ:state-equ}, respectively. Insofar the step size is concerned, $\Delta t$ is constrained by the CFL condition \cite{Monaghan1992smoothed}, magnitude of acceleration $\left | \textbf{f}_i \right |$, and viscous dispersion and chosen such that
\begin{equation*}
\Delta t \leq \min \{ 
0.25 \frac{h}{c},~ 
0.25 \min_i(\frac{h}{\left | \textbf{f}_i \right |})^{\frac{1}{2}},~
0.125 \frac{h^{2}}{\nu}
\} \;.
\end{equation*}
Despite working in a multi-resolution setup, no additional constraints were necessary on selecting $\Delta t$. Moreover, given that $Re \sim \mathcal{O} (1)$ or $Re \ll 1$ for the numerical experiments considered herein, the most stringent constraint was the viscous dispersion, i.e., $\Delta t \leq 0.125 \frac{h^{2}}{\nu}$.

\section{Simulation results}\label{sec:Simu_results}
Four different flows were used to assess the accuracy, convergence, and efficiency of the proposed multi-resolution SPH method. In this exercise, the numerical solutions were compared to the analytical solution, or numerical solution obtained using the finite element method (FEM), or numerical solution obtained using the consistent, single, and high-resolution SPH approach. The fluid was assumed to be water with $\rho = 10^3 kg/m^3$ and kinetic viscosity $\nu = 10^{-6} m^2/s$. In Sections \ref{subsec:Poi_Flow} and \ref{subsec:Flow_Cyl}, a different kinetic viscosity was used, i.e., $\nu = 10^{-4} m^2/s$, for comparing with results reported in literature. At $t=0$, the fluid was at rest; the SPH particles were distributed on a regular lattice throughout the computational domain.

\subsection{Transient Poiseuille flow}\label{subsec:Poi_Flow}  

The 2D transient Poiseuille flow took place in a straight channel with two fixed walls oriented along the $x$ axis and positioned at $y=0~m$ and $y=0.2~m$. The flow was subject to the no-slip boundary conditions at the walls and periodic boundary conditions at the remaining boundaries. The flow was driven by a body force $|\textbf{f}_{body}| = 2 \times 10^{-4}~m/s^2$ acting along the $x$ axis.

\begin{figure}[H]
\centering
\includegraphics[scale=0.60]{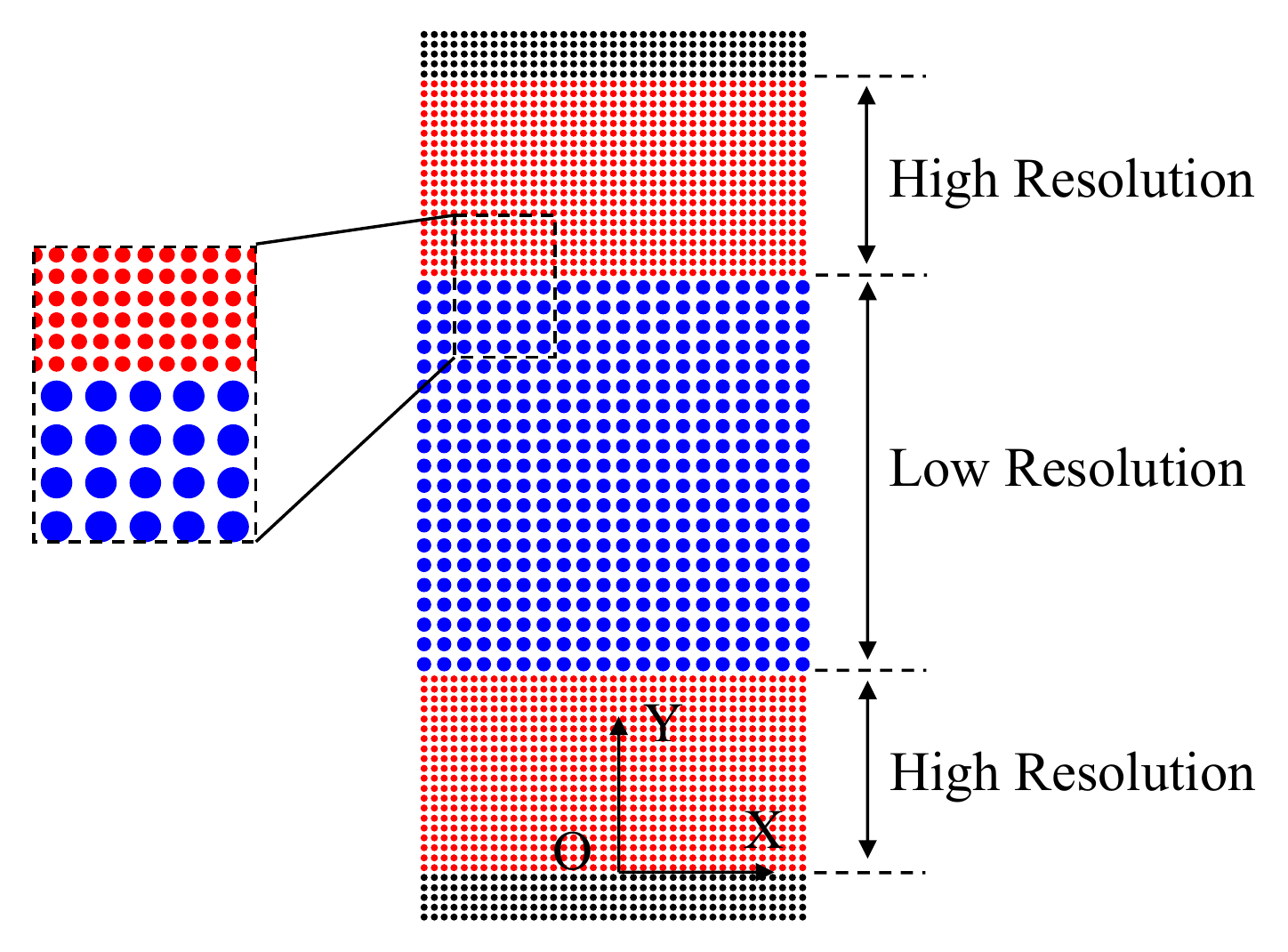}
\caption{Initial distribution of SPH particles with two different resolutions for modeling the transient Poiseuille flow.}
\label{fig:Case1-Ini-Dis}
\end{figure}

The simulation domain was partitioned into three regions across the channel width: two high-resolution regions near the walls and one low-resolution region in the middle, see Figure \ref{fig:Case1-Ini-Dis}. The width of the high-resolution and low-resolution regions is $0.05~m$ and $0.1~m$, respectively. The low resolution $\Delta x_L$ was fixed as $5~mm$. In a series of three simulations, the high resolution $\Delta x_H$ was varied from $5~mm$ to $1.25~mm$ to produce ratios $\Upsilon_{LH}$ of 1, 2, and 4. 

\begin{figure}[H]
\centering
\subfigure[$\Delta x_{H}=2.5~mm$; $\Upsilon_{LH} = 2$.]{
\includegraphics[scale=0.33]{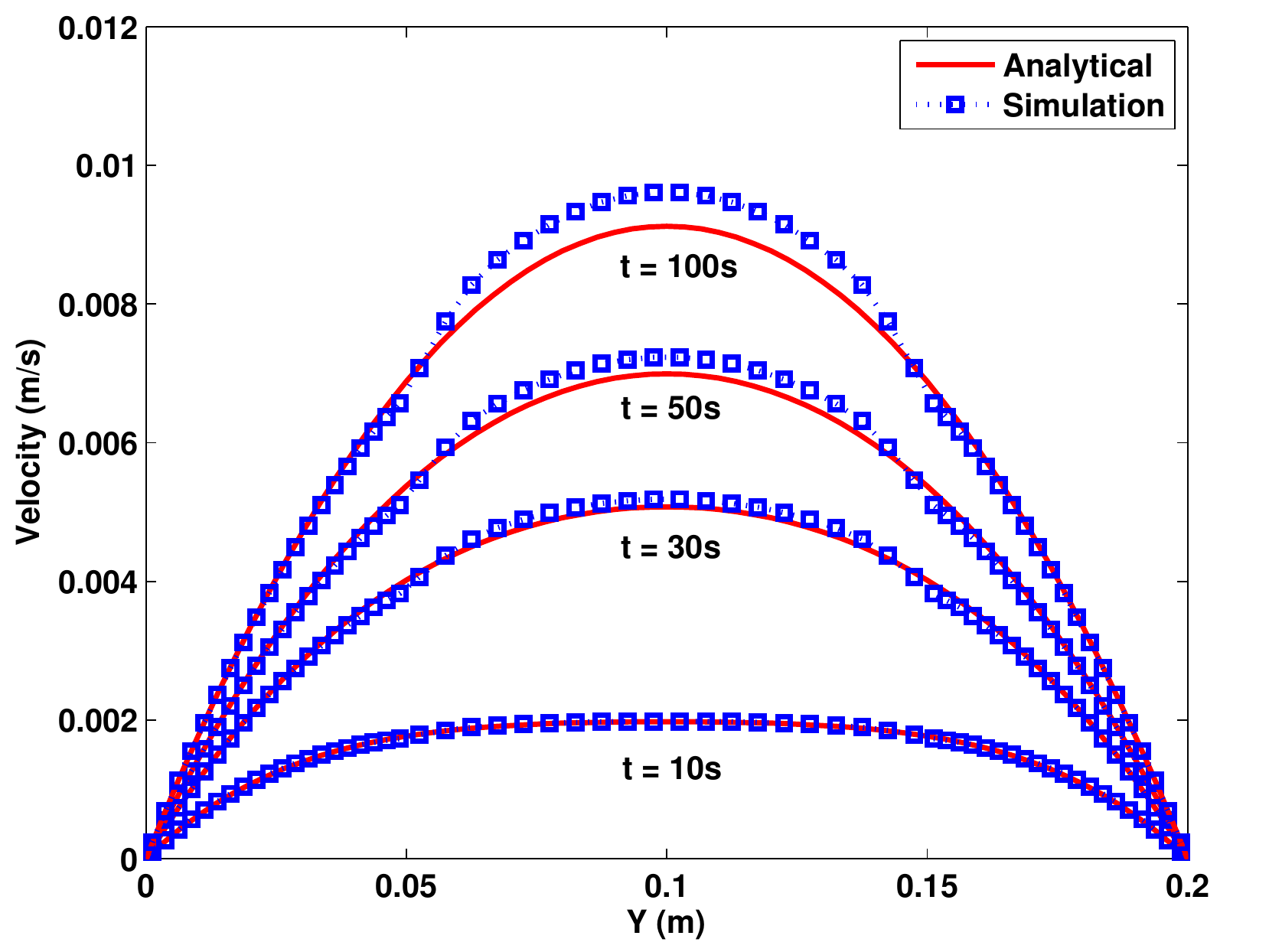}}
\subfigure[$\Delta x_{H}=1.25~mm$; $\Upsilon_{LH} = 4$.]{
\includegraphics[scale=0.33]{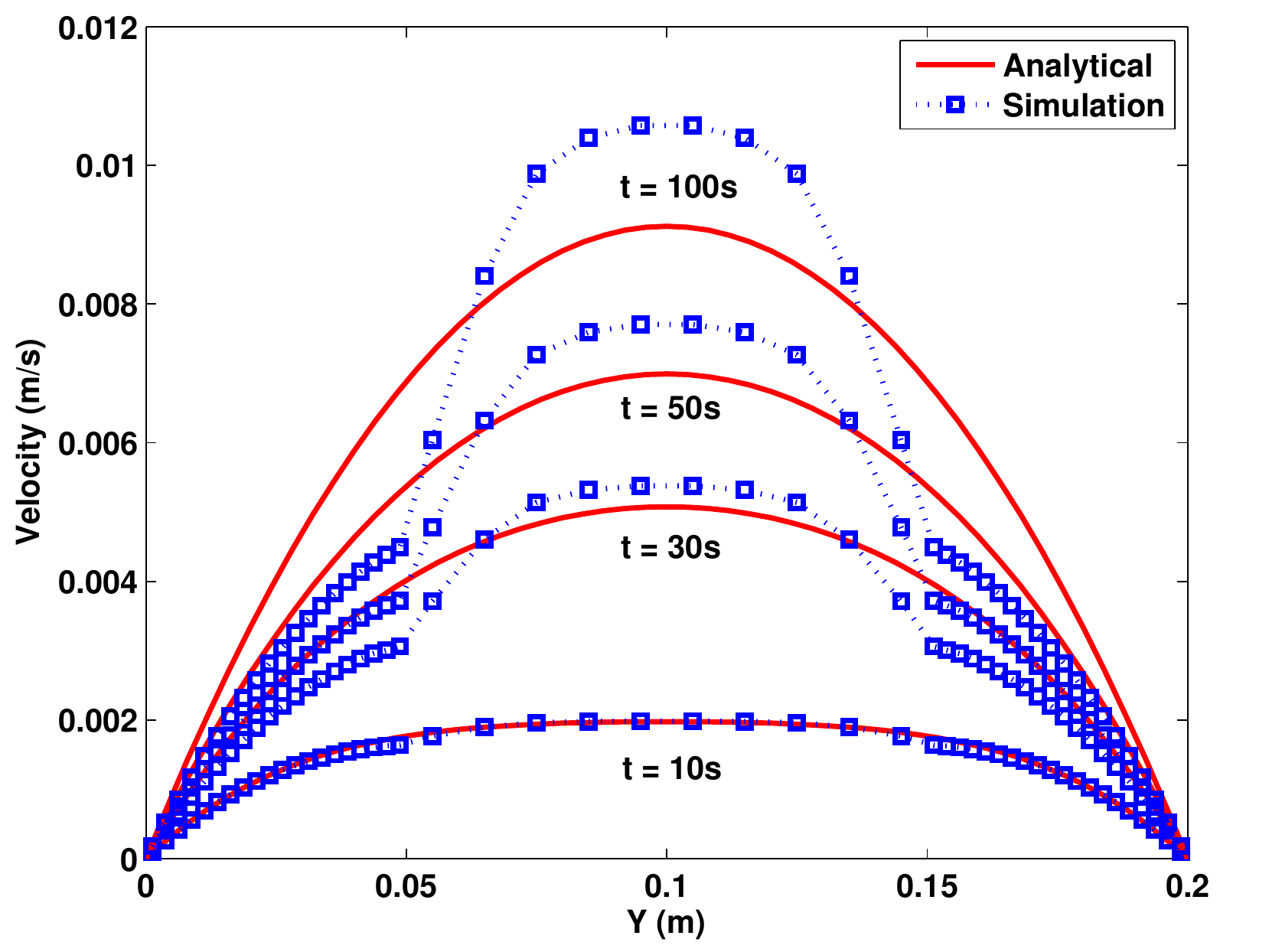}}
\caption{Transient velocity profiles of Poiseuille flow across the channel width obtained with the standard SPH at the spatial resolution of $\Delta x_{L}=5~mm$ and different $\Delta x_H$, compared to the analytical solution.}
\label{fig:Case1-Vel-Prof-Bad}
\end{figure}

\begin{figure}[H]
\centering
\subfigure[$\Delta x_{H}=5mm$; $\Upsilon_{LH}=1$.]{
\includegraphics[scale=0.33]{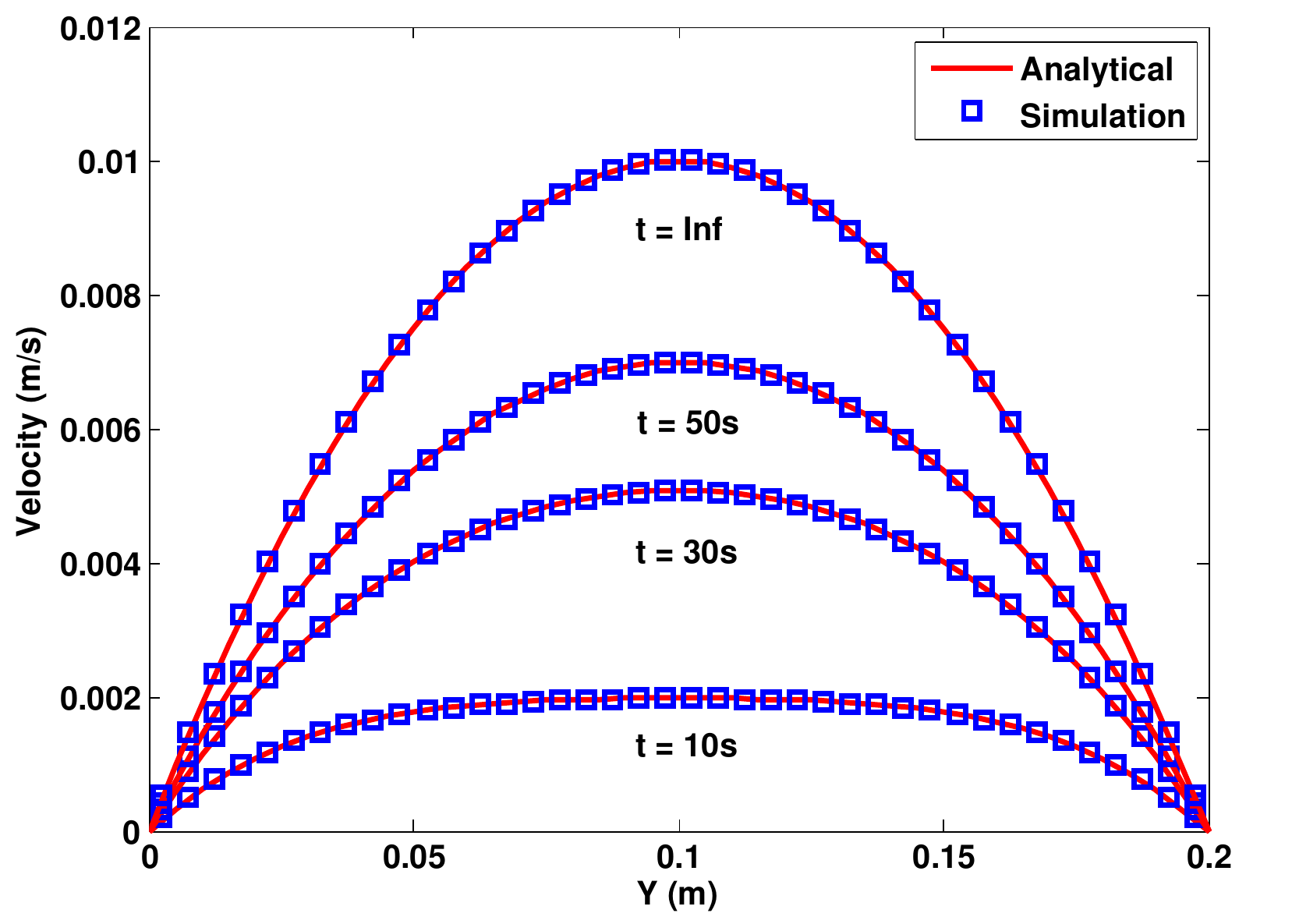}}
\subfigure[$\Delta x_{H}=2.5mm$; $\Upsilon_{LH}=2$.]{
\includegraphics[scale=0.33]{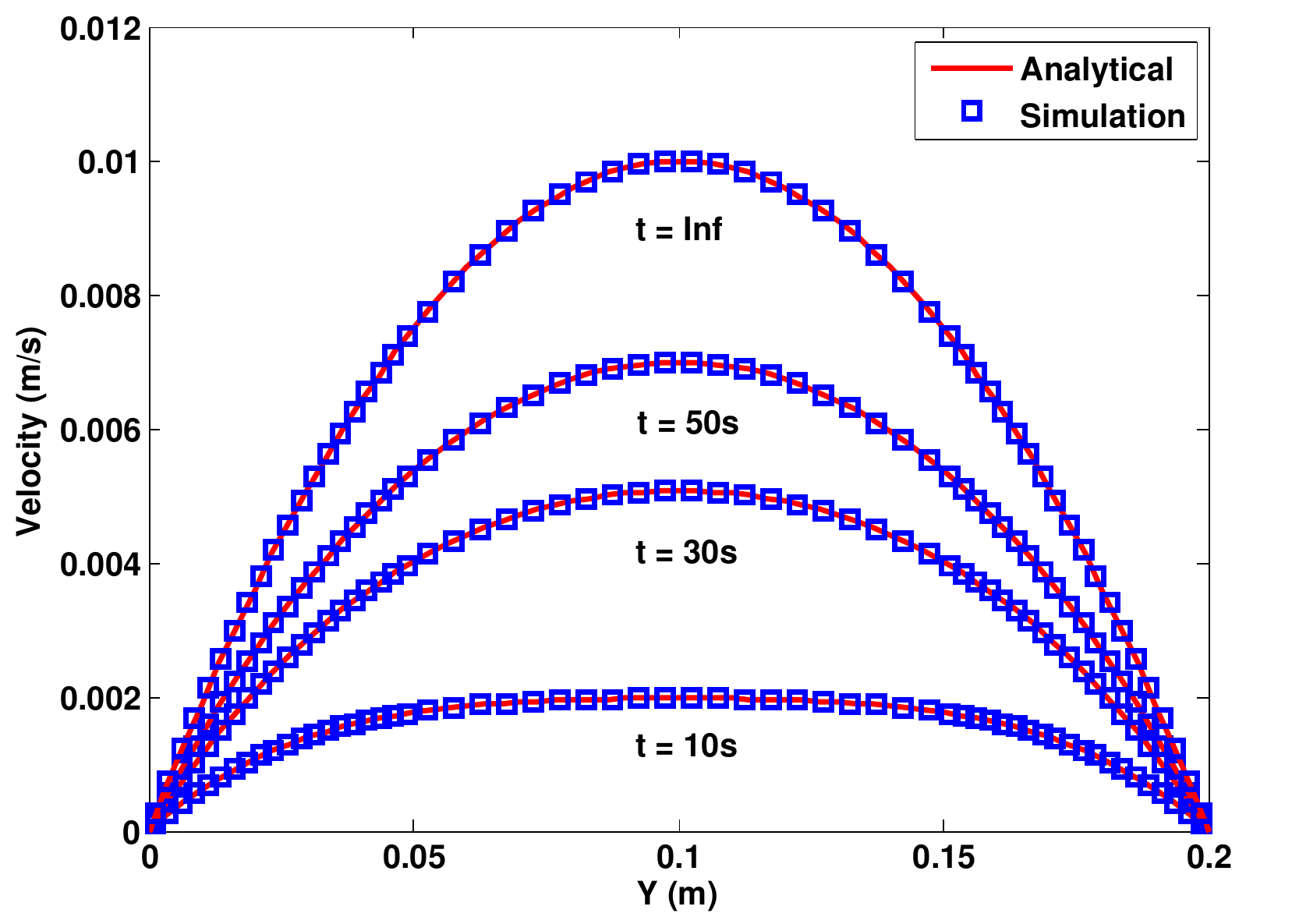}}
\subfigure[$\Delta x_{H}=1.25mm$; $\Upsilon_{LH}=4$.]{
\includegraphics[scale=0.33]{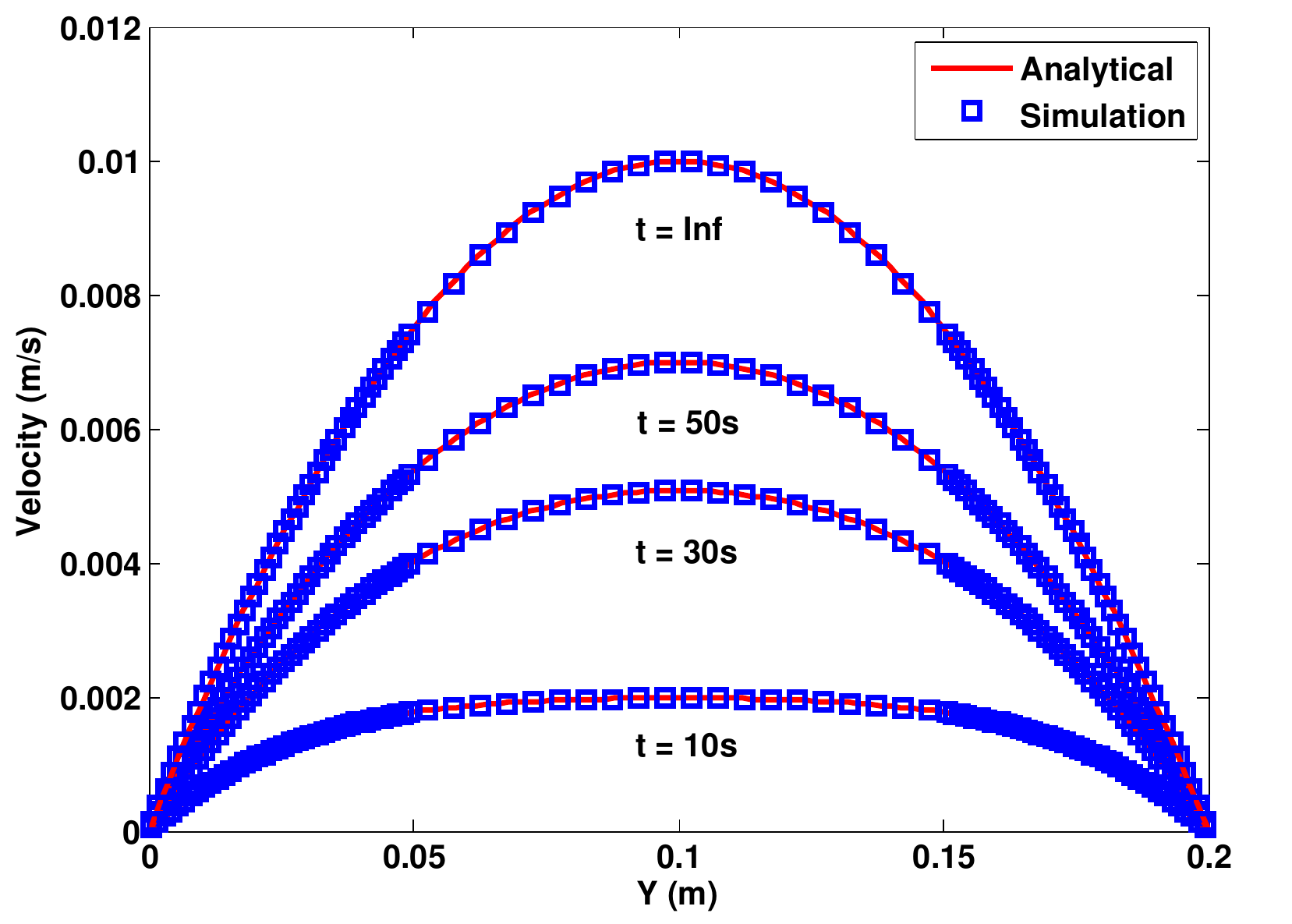}}
\caption{Transient velocity profiles of Poiseuille flow across the channel width computed by the consistent multi-resolution SPH for $\Upsilon_{LH}=1,~2,~4$. The analytical solution is shown for reference.}
\label{fig:Case1-Vel-Prof}
\end{figure}

Figure \ref{fig:Case1-Vel-Prof-Bad} shows the transient velocity profiles across the channel width computed at different times using the standard SPH discretization of Eq. \eqref{equ:dis_inconsistent}. The SPH results deteriorate as time goes on; and, the discrepancy becomes worse as $\Upsilon_{LH}$ increases. This suggests that two different resolutions cannot be directly coupled using the standard SPH discretization. 
Figure \ref{fig:Case1-Vel-Prof} shows the transient velocity profiles obtained via the consistent SPH discretization of Eqs. \eqref{equ:gra_ope}--\eqref{equ:density_renor}. The proposed approach accurately predicted the velocity profiles for ratios of $\Upsilon_{LH} =$1, 2, and 4. No numerical artifact was noticed. Thus, the proposed SPH discretization allows for coupling of different resolutions without using an overlap region.

\begin{figure}[H]
\centering
\includegraphics[scale=0.55]{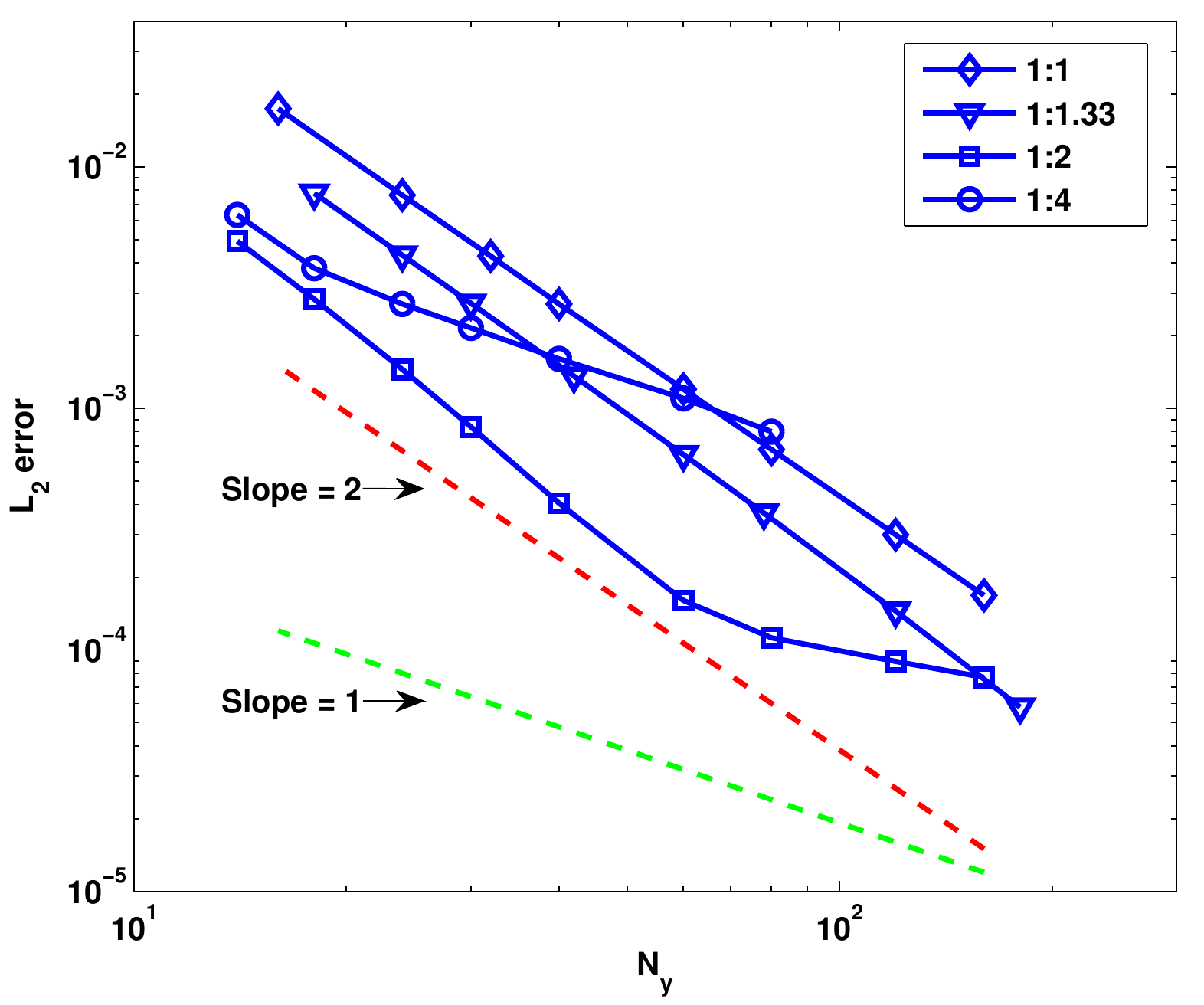}
\caption{Convergence of the numerical solutions on the stead-state velocity in Poiseuille flow for different $\Upsilon_{LH}$. The $x$ axis displays the number of particles per unit length across the channel width in the low-resolution region; the $y$ axis is the $L_{2}$ error relative to the analytical solution.}
\label{fig:Case1-Convergence}
\end{figure}

We next probed the convergence attribute of the multi-resolution SPH method. To that end, we varied the number of particles at each $\Upsilon_{LH}$ and monitored the relative $L_2$ error of the steady-state velocity. Figure \ref{fig:Case1-Convergence} summarizes our findings. The convergence of the multi-resolution SPH is affected by $\Upsilon_{LH}$. For small $\Upsilon_{LH}$, the second-order convergence of SPH can be sustained. As it increases, the anisotropy of particle configuration increases. Thus, the second-order convergence degrades to the first-order as $\Upsilon_{LH}$ assumes values close to 2 or higher. The larger $\Upsilon_{LH}$ is, the earlier the degradation of convergence occurs. This finding about the effect of anisotropic particle configuration on the convergence of SPH is consistent with results reported in the previous studies \cite{Fatehi2011error,Traska2015second-order}. Thus, on one hand, it is desirable to use larger $\Upsilon_{LH}$ as it leads to more computational savings. On the other hand, these savings may sacrifice accuracy. As an optimal cost--accuracy trade-off, we chose $\Upsilon_{LH}=2$ for all remaining simulations.

\subsection{Flow around a periodic array of fixed cylinders}\label{subsec:Flow_Cyl}  

In this test, a cylinder of radius $0.02~m$ was positioned at the center of a square domain of length $0.1~m$. The flow was driven by a body force $|\textbf{f}_{body}| = 5 \times 10^{-5}~m/s^2$ along the $x$ coordinate, and subject to the no-slip boundary condition at the cylinder and periodic boundary conditions at the remaining boundaries. To examine the long-time particle regularity and numerical stability, the simulation was run for more than $3000~s$. This flow reached its steady-state at about $25~s$.

\begin{figure}[H]
\centering
\includegraphics[scale=0.6]{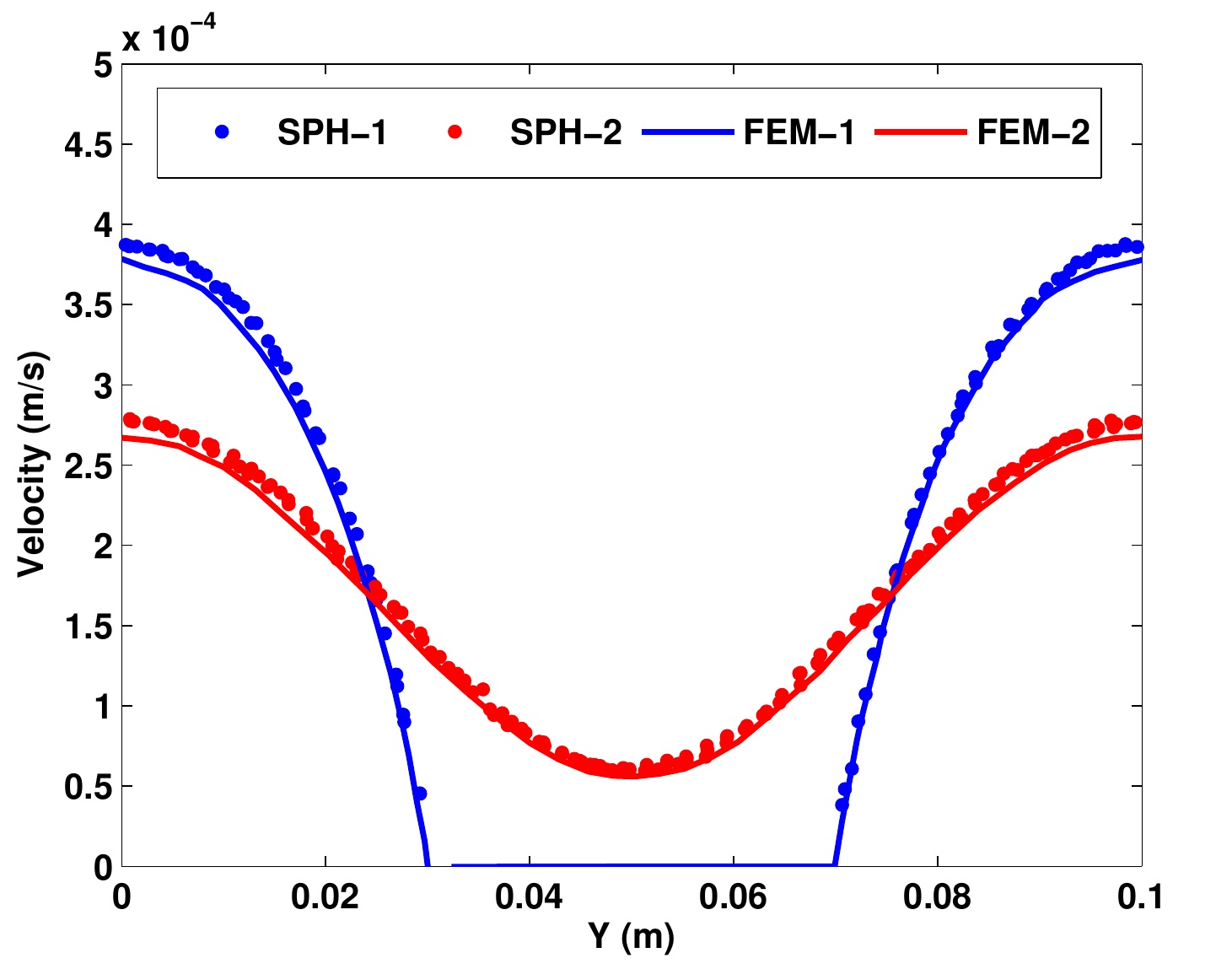}
\caption{Comparison of SPH and FEM velocity profiles along lines 1 ($x=0.05~m$) and 2 ($x=0.1~m$).}
\label{fig:Case2-MorrisFEM-OurSPH}
\end{figure}
\begin{figure}[H]
\centering
\subfigure[]{
\includegraphics[scale=0.7]{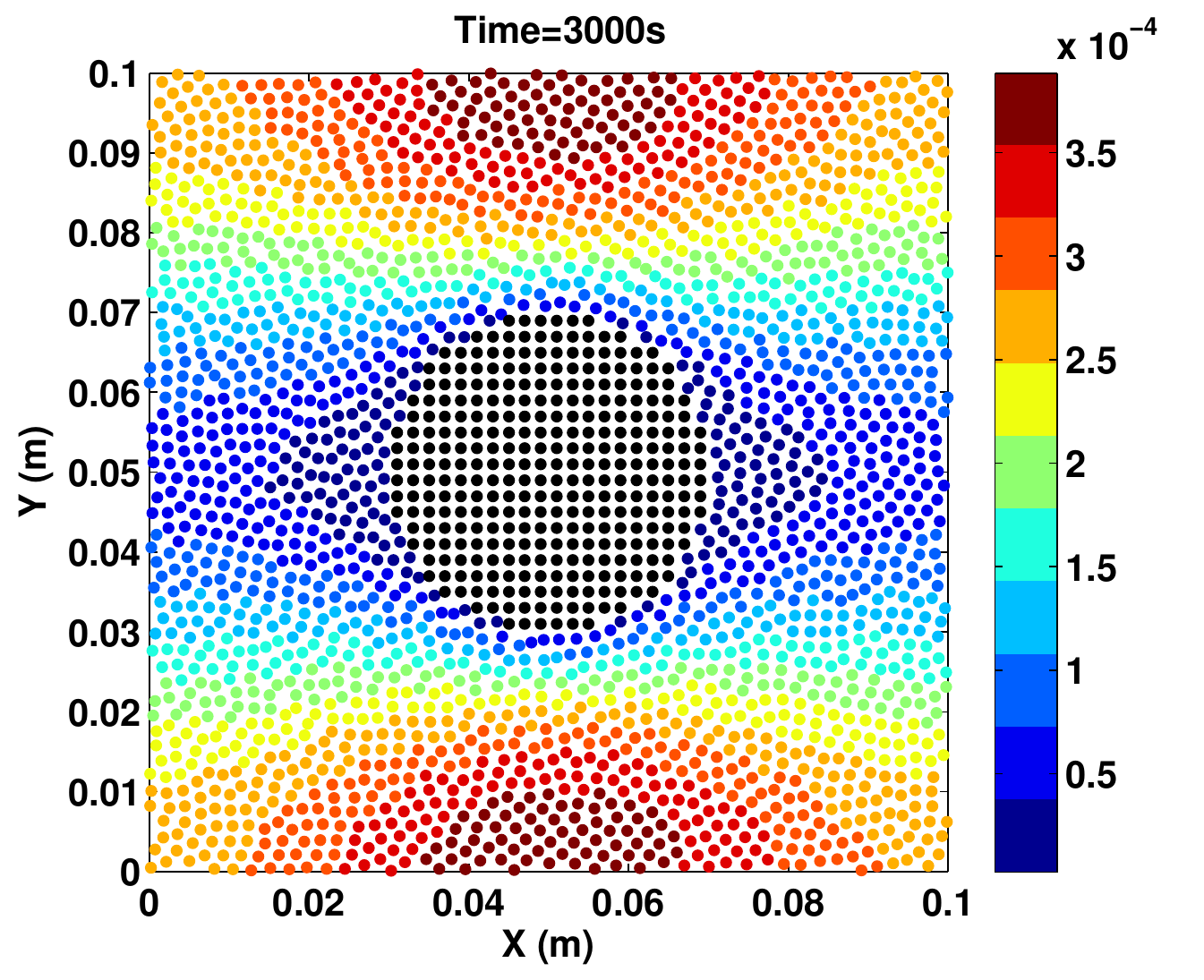}}
\subfigure[]{
\includegraphics[scale=0.7]{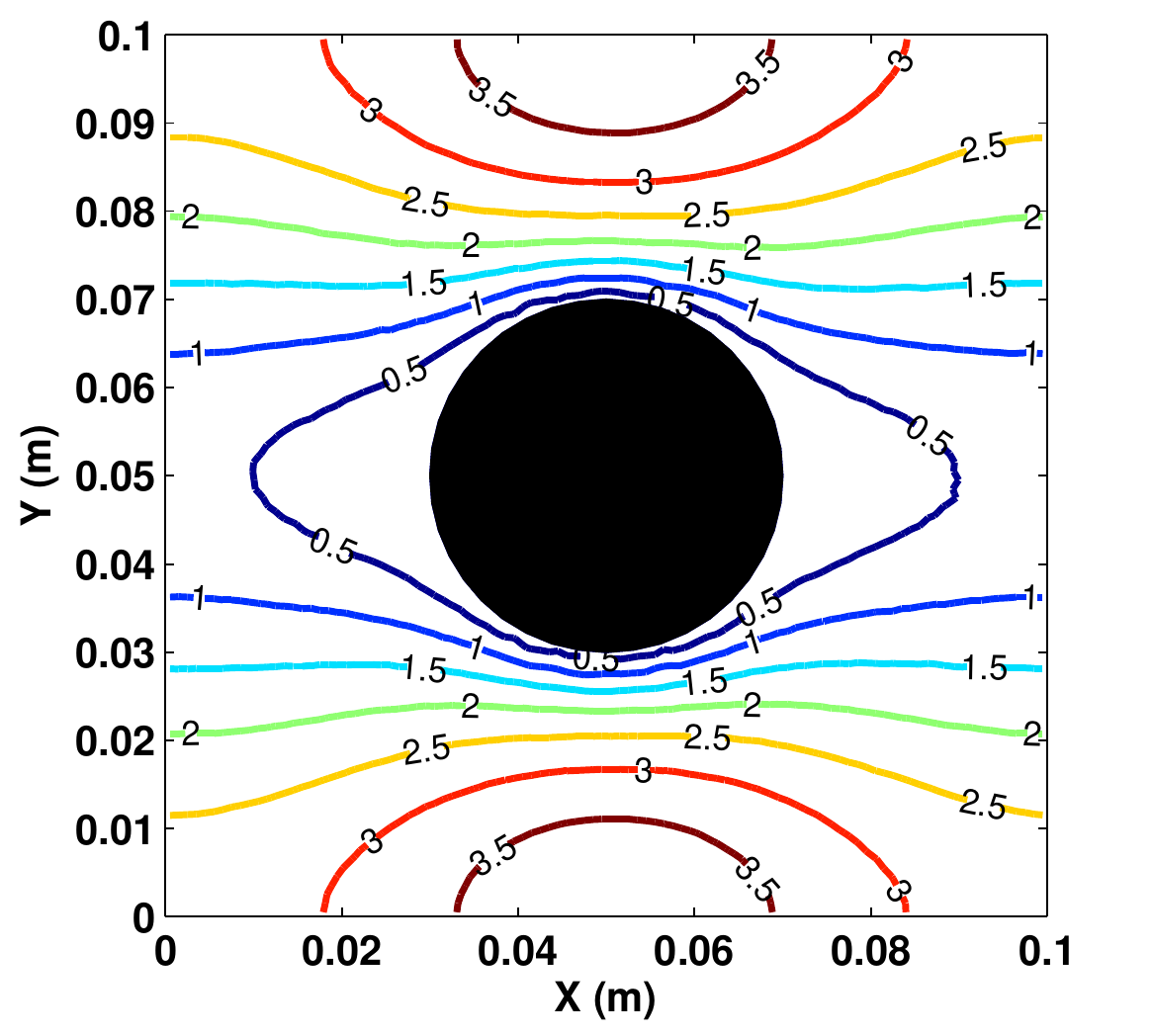}}
\caption{Consistent single-resolution SPH results -- flow around a periodic array of fixed cylinders. (a) Particle distribution with particle color correlated to velocity magnitude at $t=3000s$ (units in the color bar: $m/s$); (b) Equi-magnitude contours of averaged steady-state velocity (unit: $\times 10^{-4}~m/s$).}
\label{fig:Case2-Final-Dis-Morris}
\end{figure}

We first examined the numerical accuracy of the SPH solution when $\Delta x=2~mm$ and $\Upsilon_{LH}=1$; i.e., in the single resolution case. To that end, we sliced the domain vertically with imaginary lines at $x=0.05~m$ and at $x=0.1~m$. Figure \ref{fig:Case2-MorrisFEM-OurSPH} compares the SPH results along lines 1 and 2 to those predicted by FEM \cite{Morris1997modeling}. The agreement between the two numerical solutions is good, with a $L_\infty$ relative error of 3.5\%. Figure \ref{fig:Case2-Final-Dis-Morris} pertains to the long-term response. Subfigure (a) shows the particles distribution at $t = 3000~s$, which demonstrates the particle regularity is well maintained in a long-time simulation. Subfigure (b) shows the equi-magnitude contours of steady-state velocity averaged after $t=500~s$.

\begin{figure}[H]
\centering
\includegraphics[scale=0.7]{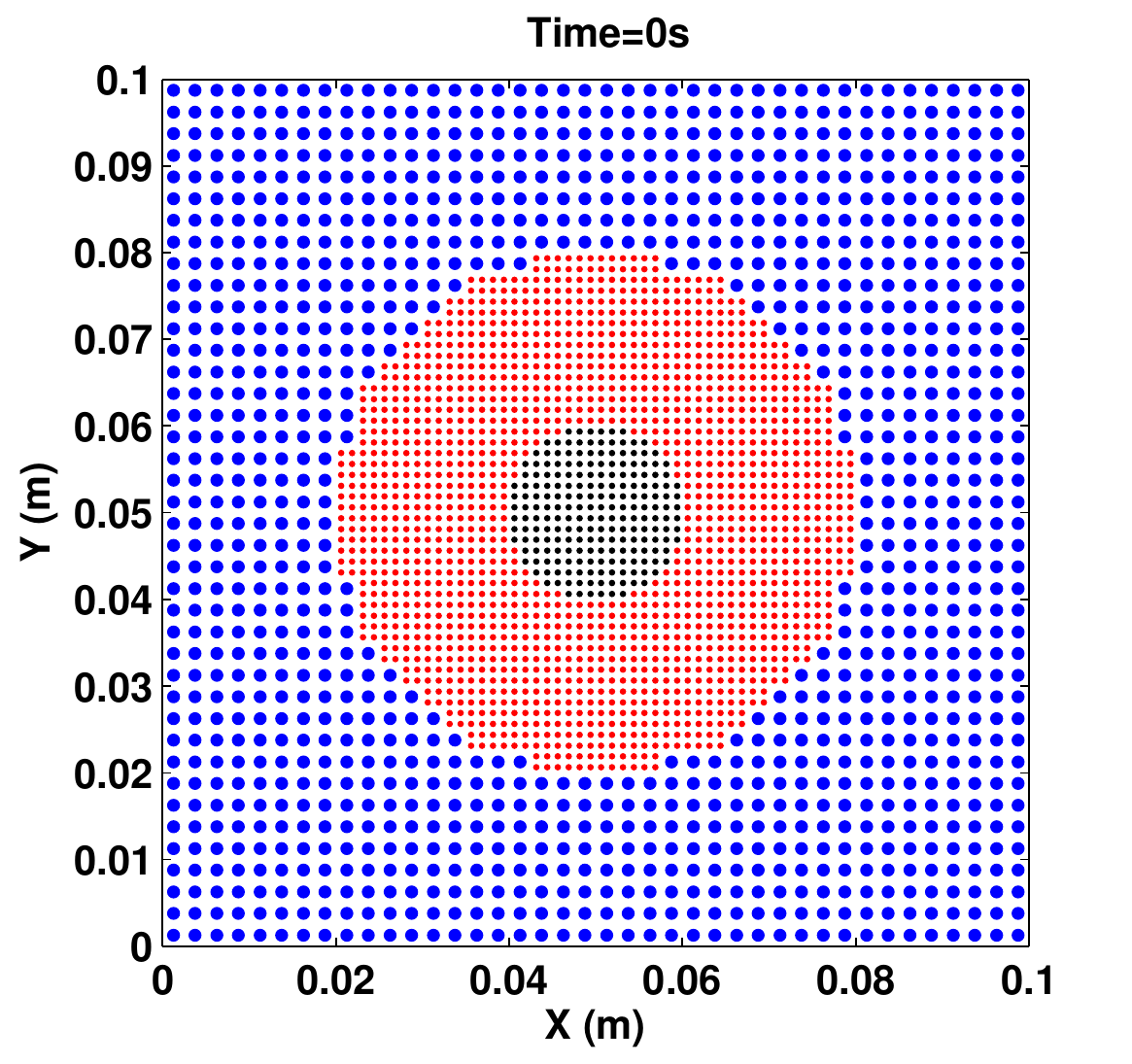}
\caption{Initial particle configuration for the dual-resolution SPH simulation of the flow around a periodic array of fixed cylinders.}
\label{fig:Case2-Ini-Dis-OurSPH}
\end{figure}

\begin{figure}[H]
\centering
\subfigure[Single resolution.]{
\includegraphics[scale=0.65]{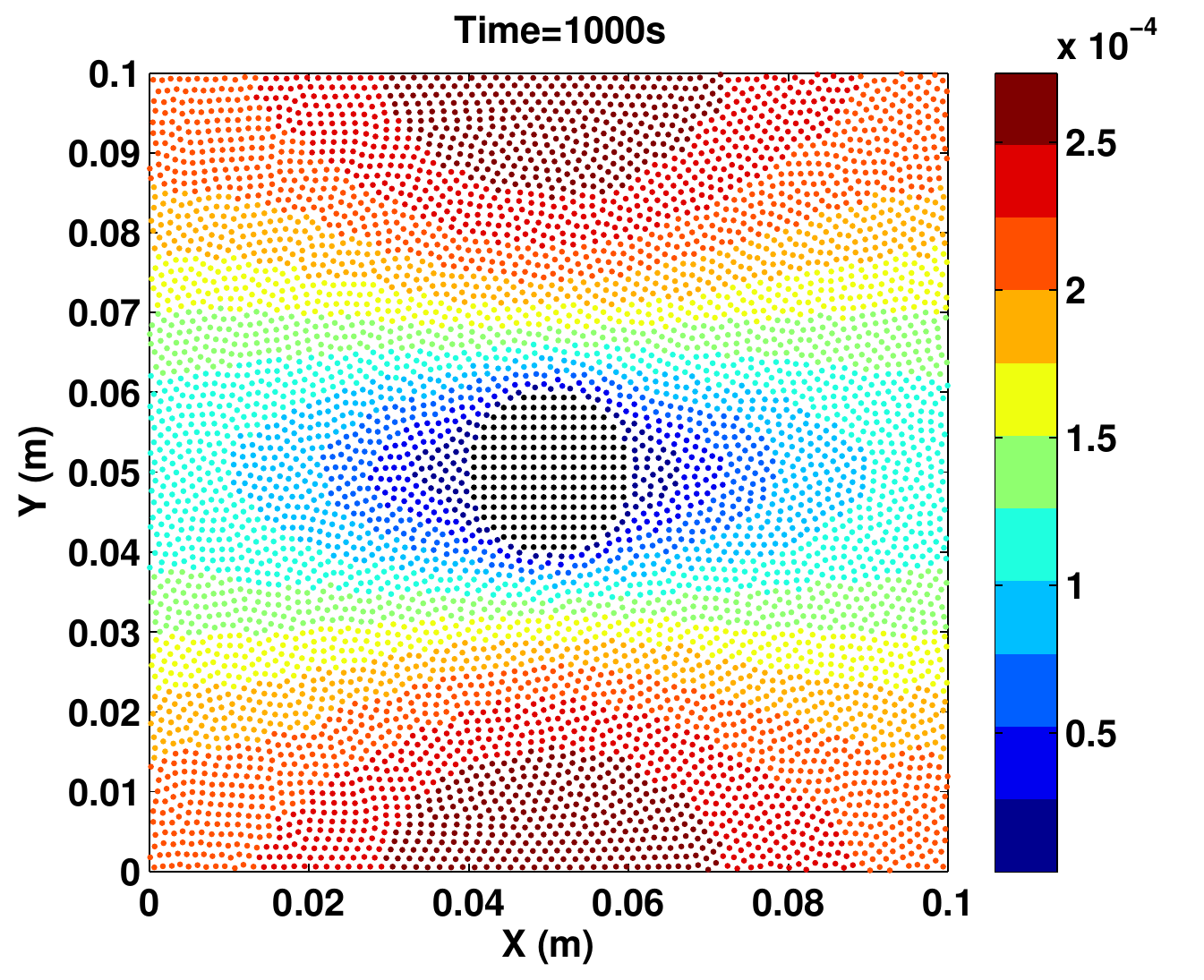}}
\subfigure[Dual resolutions.]{
\includegraphics[scale=0.65]{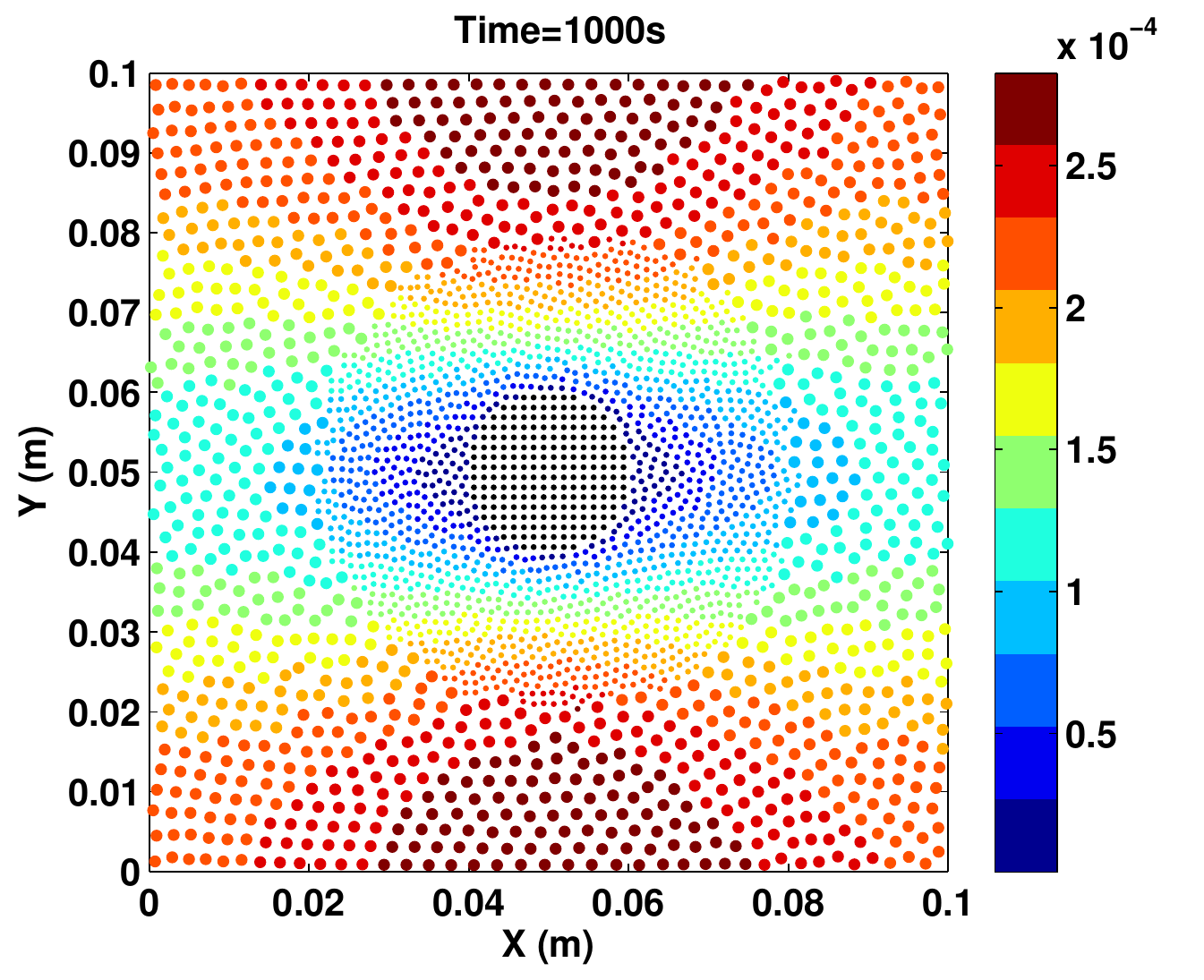}}
\caption{Particle distribution with color correlated to velocity magnitude at $t=1000s$ (unit in color bar: $m/s$).}
\label{fig:Case2-Final-Dis-OurSPH}
\end{figure}

\begin{figure}[H]
\centering
\subfigure[Single resolution.]{
\includegraphics[scale=0.65]{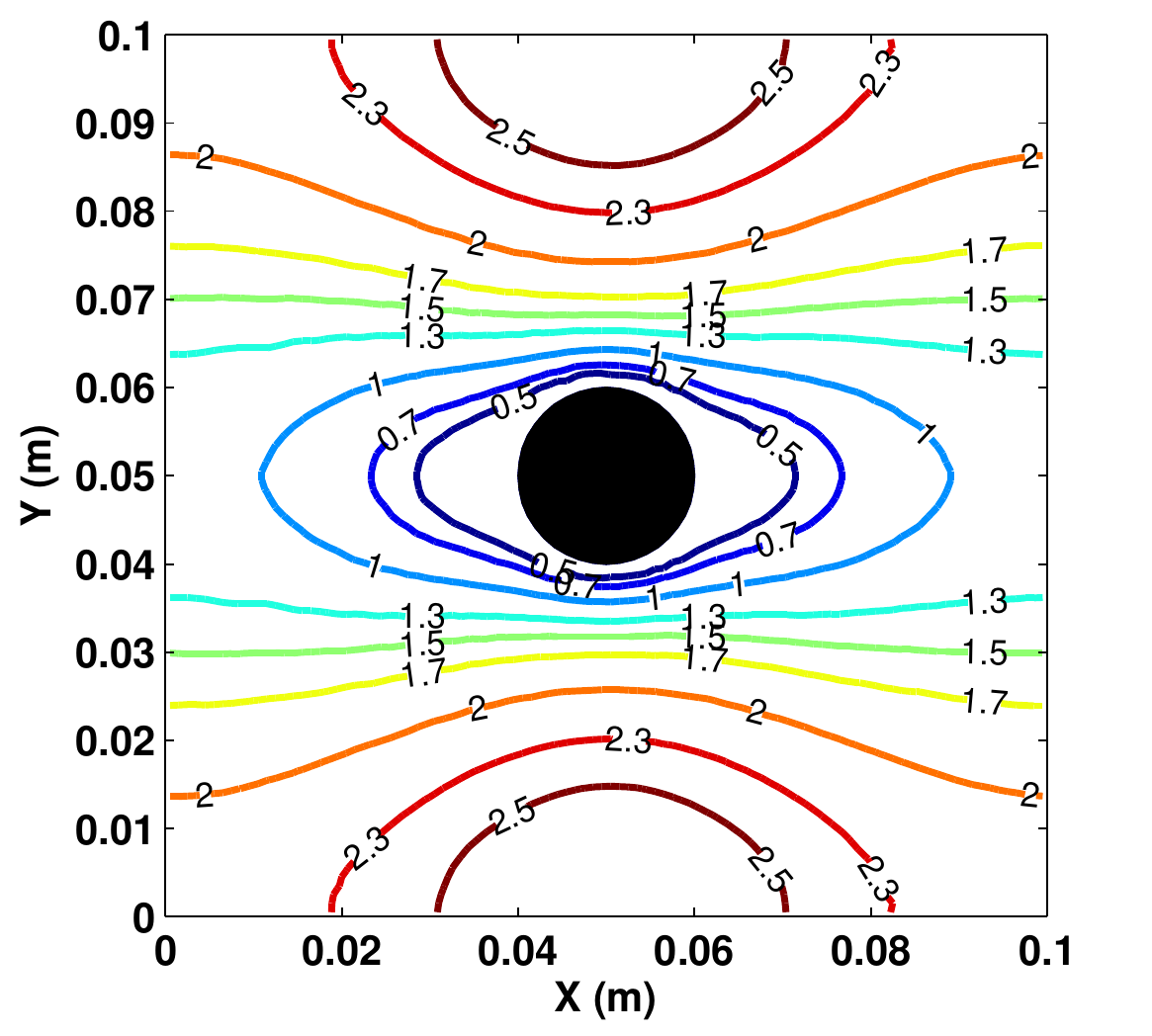}}
\subfigure[Dual resolutions.]{
\includegraphics[scale=0.65]{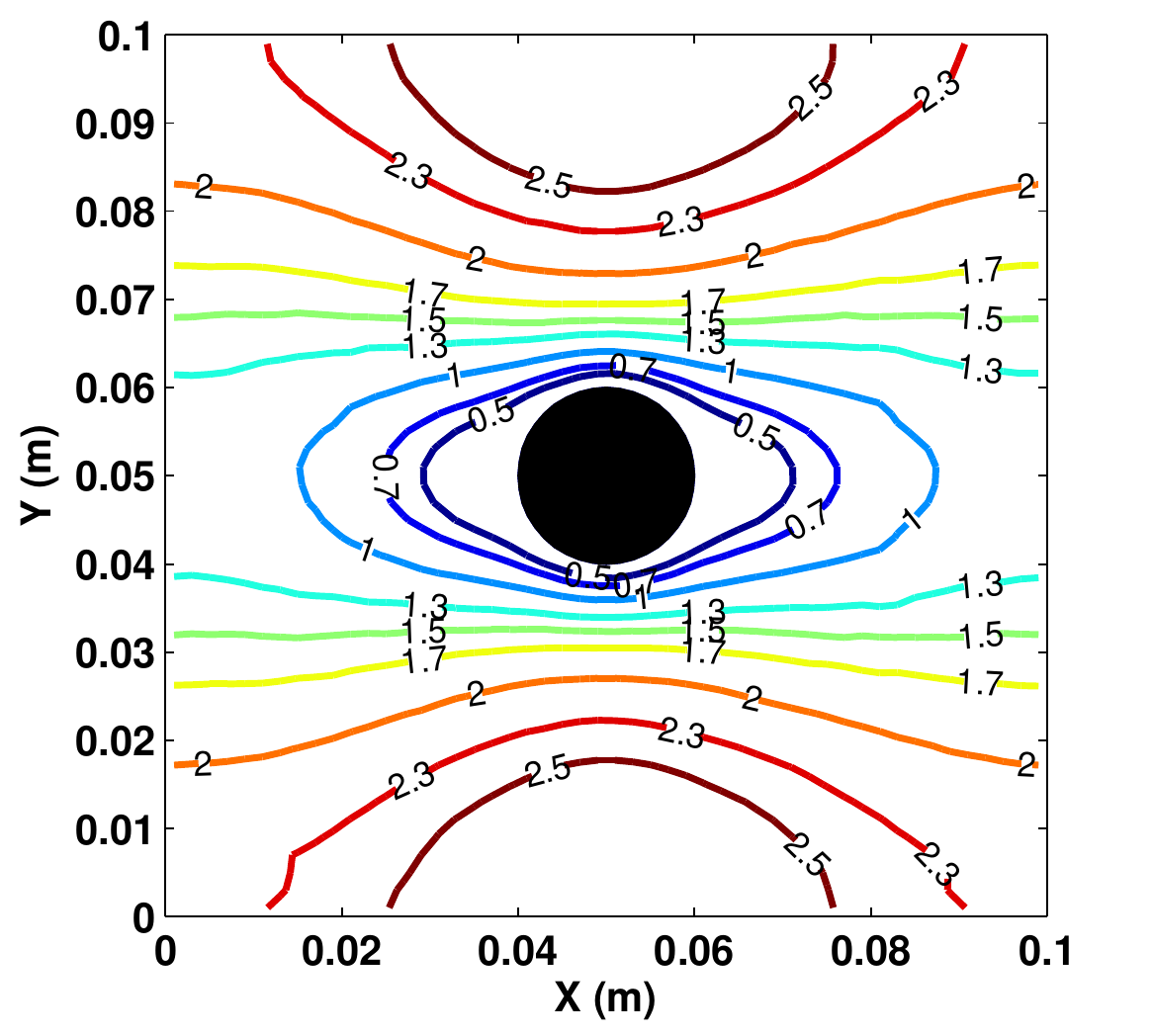}}
\caption{Equi-magnitude contours of averaged steady-state velocity for the flow around a periodic array of fixed cylinders (unit: $10^{-4}$~m/s).}
\label{fig:Case2-Velocity-Contour-OurSPH}
\end{figure}

Next, we employed the multi-resolution SPH to simulate the same flow. As illustrated in Figure \ref{fig:Case2-Ini-Dis-OurSPH}, the computational domain was divided into two regions. The inner, high-resolution, and circular region was located next to the cylinder and had a thickness of $0.02~m$. The remaining fluid domain was represented in low resolution. The cylinder and fluid inside the circular region were discretized using $\Delta x_H=1.25~mm$; the low-resolution representation used $\Delta x_L=2.5~mm$. For comparison, we also performed a single high-resolution simulation with $\Delta x=\Delta x_H=1.25~mm$. 
Figure \ref{fig:Case2-Final-Dis-OurSPH} presents the snapshots at $t=1000~s$ for the single- and dual-resolution simulations, respectively. Shown are the particle distributions with color correlated to velocity magnitude. Figure \ref{fig:Case2-Velocity-Contour-OurSPH} further compares the velocity fields in the single- and dual-resolution simulations via equi-magnitude contours of steady-state velocity averaged after $t=500s$. These figures confirm that the multi-resolution predictions agree very well with those of the single-resolution SPH in the entire fluid domain. Moreover, when comparing the total drag force exerted on the cylinder, the difference is less than 6\%. Note that 2943 particles were used in the multi-resolution SPH versus 6400 particles in the single-resolution SPH, a 54\% reduction in degrees of freedom. For each 10-$s$ (physical time) simulation, the computational time of the single-resolution solution was 1104 $s$, but only 695 $s$ for the multi-resolution solution.
Finally, we investigated the effect of the thickness of high-resolution region. For that, we performed simulations with different thicknesses of high-resolution regions. The total drag on the cylinder was then computed and compared to that obtained from the single high-resolution simulation. We found that enlarging the high-resolution region with 17\% more particles would lead to about 80\% improvement on the accuracy and less than 20\% increase in computational time.

\subsection{Flow around a rotating cylinder near a moving wall}\label{subsec:Flow_Wall}  
\begin{figure}[H]
\centering
\includegraphics[scale=0.65]{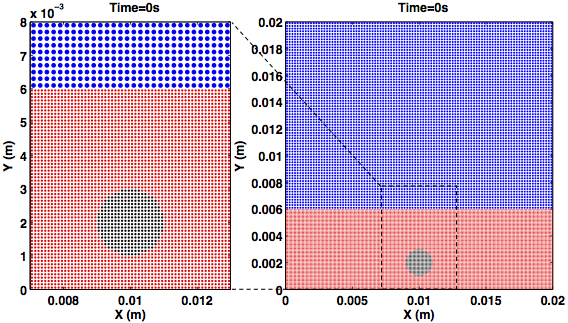}
\caption{Initial particle configuration with two resolutions for simulating the flow around a rotating cylinder near a moving wall.}
\label{fig:Case3-Ini-Dis-OurSPH}
\end{figure}

The computational domain in this test was a square box of length $0.02~m$. The center of the cylinder was positioned at $x=0.01~m$ and $y=0.002~m$; its radius was $R = 0.001~m$. The cylinder was rotating counter-clockwise at a constant angular velocity $\omega=9.46\times10^{-2}s^{-1}$. The bottom wall was moving horizontally along $x$ at a constant velocity of $\omega R$ from left to right; the top wall was fixed. No-slip boundary conditions were imposed at the cylinder and walls; periodic boundary conditions were applied at the remaining boundaries. We divided the computational domain at $y=0.06~m$ into two regions: the bottom high-resolution region with $\Delta x_H=0.1~mm$; the remaining fluid domain discretized at a lower resolution $\Delta x_L=0.2~mm$. Figure \ref{fig:Case3-Ini-Dis-OurSPH} shows the two regions in the initial configuration. 

\begin{figure}[H]
\centering
\subfigure[Single resolution.]{
\includegraphics[scale=0.65]{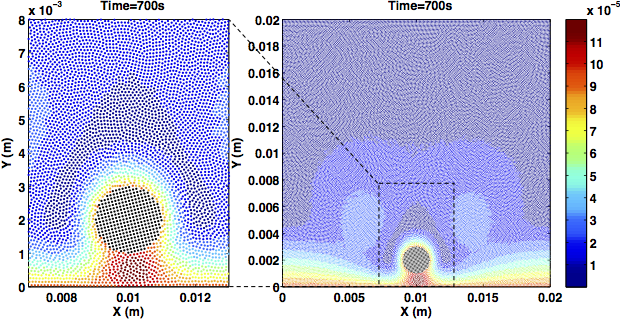}}
\subfigure[Dual resolutions.]{
\includegraphics[scale=0.65]{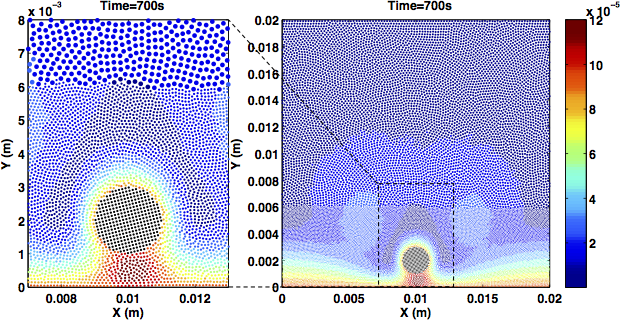}}
\caption{Particle distribution at $t=700~s$ for the flow around a rotating cylinder near a moving wall. Color contour is correlated to the fluid velocity magnitude (unit in the color bar: $m/s$).}
\label{fig:Case3-Final-Dis-OurSPH}
\end{figure}

\begin{figure}[H]
\centering
\subfigure[Single resolution.]{
\includegraphics[scale=0.4]{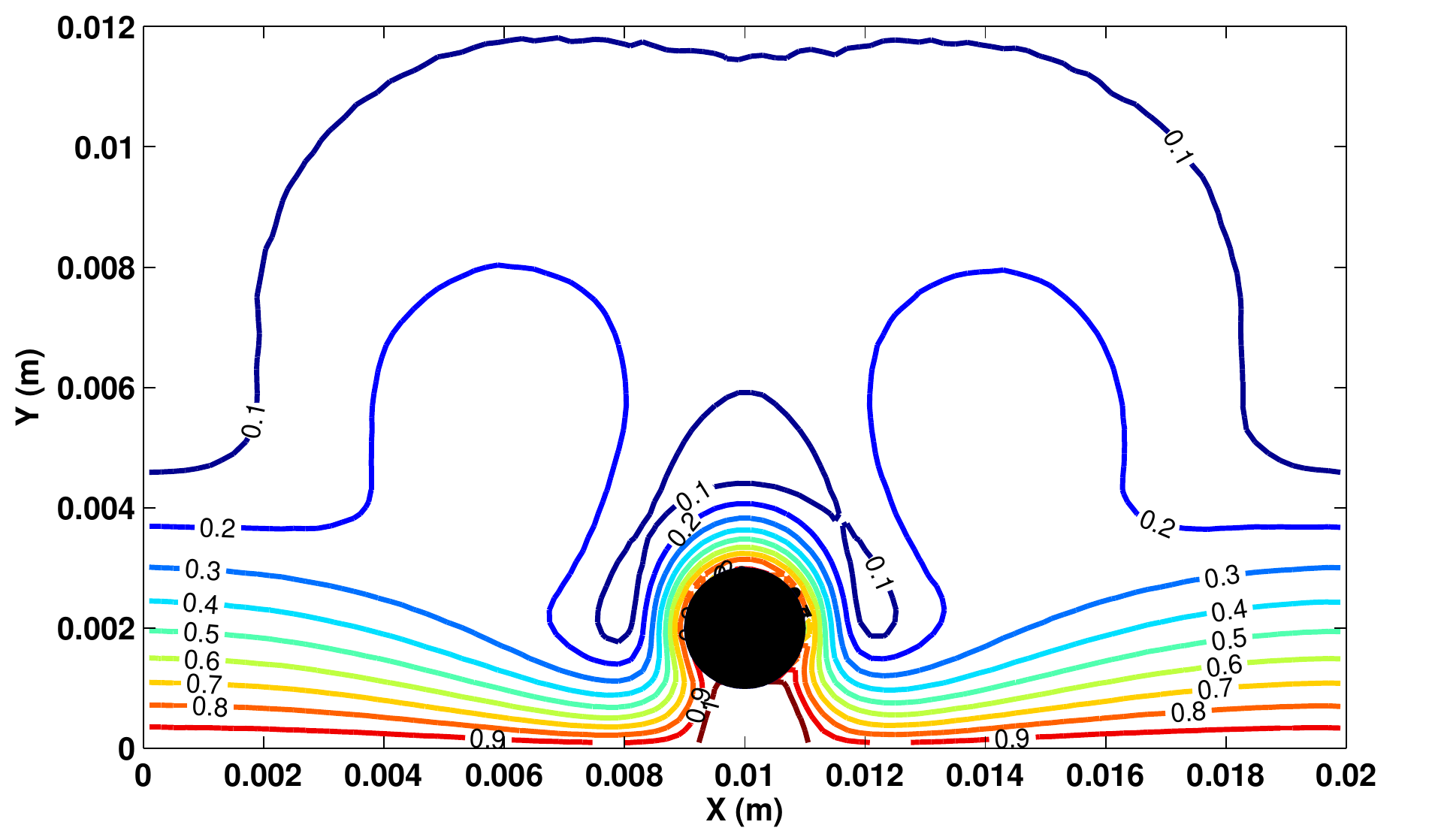}}
\subfigure[Dual resolutions.]{
\includegraphics[scale=0.4]{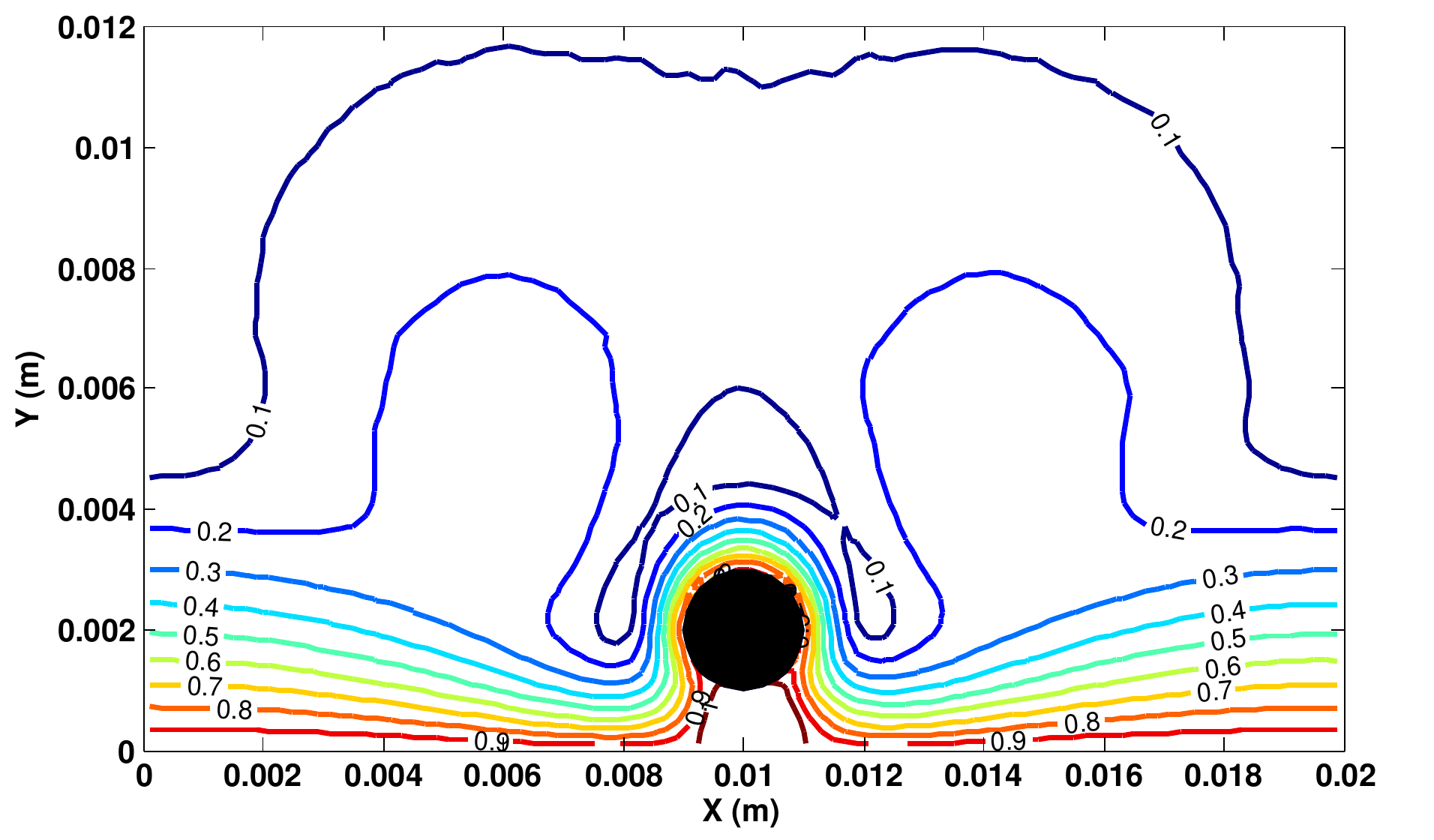}}
\caption{Equi-magnitude contours of averaged steady-state velocity for the flow around a rotating cylinder near a moving wall (unit: $9.46\times10^{-5}~m/s$).}
\label{fig:Case3-Velocity-Contour-OurSPH}
\end{figure}

\begin{figure}[H]
\centering
\subfigure[$x=0.005m$]{
\includegraphics[scale=0.33]{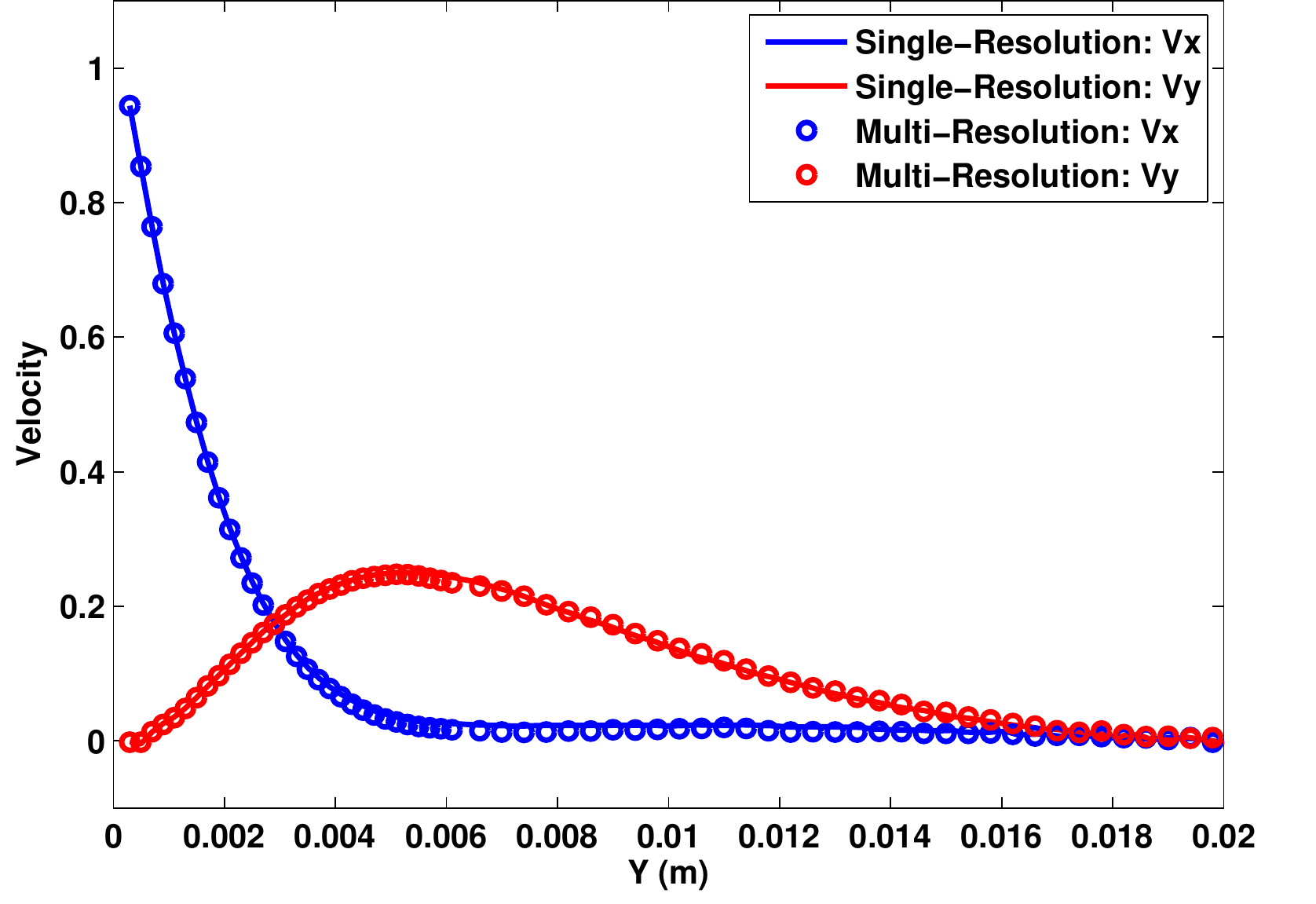}}
\subfigure[$x=0.01m$]{
\includegraphics[scale=0.33]{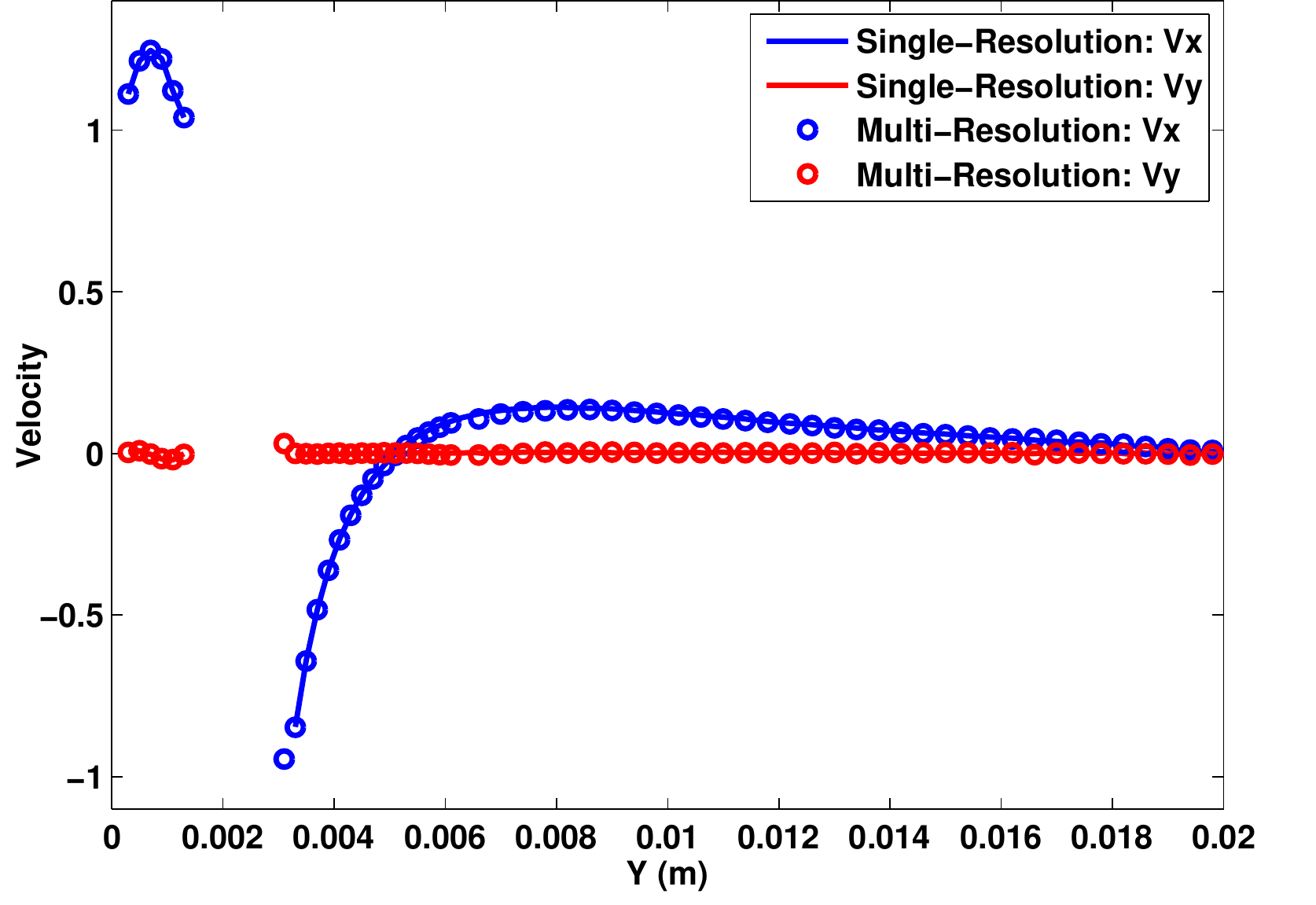}}
\subfigure[$x=0.015m$]{
\includegraphics[scale=0.33]{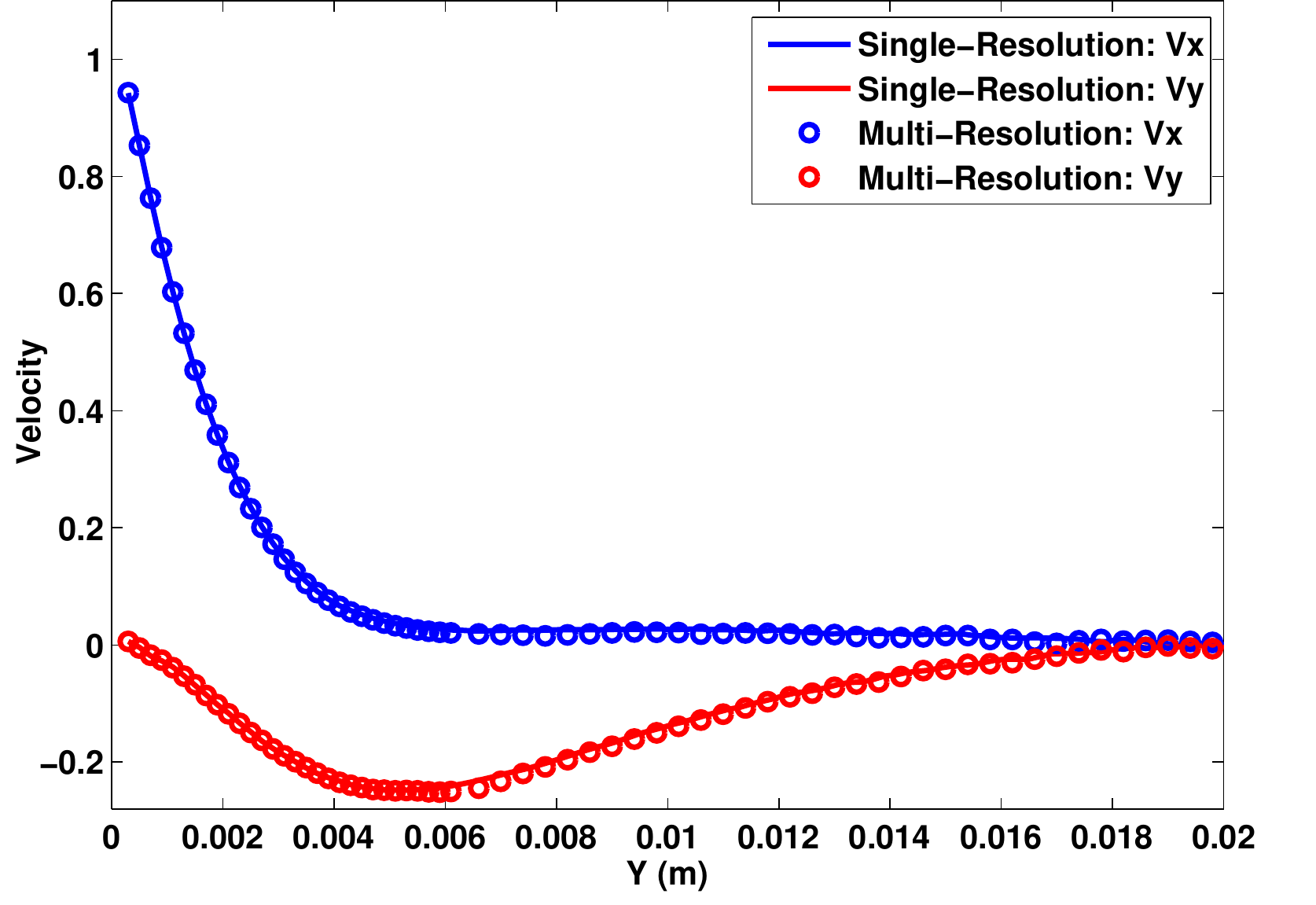}}
\caption{Velocity profiles along different lines for the flow around a rotating cylinder near a moving wall (unit: $9.46\times10^{-5}~m/s$), obtained from both the single-resolution SPH and multi-resolution SPH.}
\label{fig:Case3-Velocity-Profile-OurSPH}
\end{figure}

Figure \ref{fig:Case3-Final-Dis-OurSPH} shows a snapshot of the particle distribution: subfigure (a) pertains to the single-resolution SPH; subfigure (b) pertains to the dual-resolution case. The particle color is correlated to the fluid velocity magnitude at $t=700s$. Figure \ref{fig:Case3-Velocity-Contour-OurSPH} shows the equi-magnitude contour lines of steady-state velocity averaged after $t=500s$. Figure \ref{fig:Case3-Velocity-Profile-OurSPH} further depicts the $y-$distribution of the averaged steady-state velocities at three locations: $x=0.005~m$, $x=0.01~m$, and $x=0.015~m$. All data suggest very good agreements between single and dual-resolution results. Noteworthy, 21,000 particles were used in the dual-resolution simulation, less than a half of 42,400 particles used for the single-resolution solution.
For each 10-$s$ (physical time) simulation, the computational time of the single-resolution solution was 4168 $s$, but only 2758 $s$ for the multi-resolution solution.

Finally, Bian et al. have simulated this flow in \cite{Bian2015DDSPH} using the domain-decomposition multi-resolution SPH (DD-SPH) method. The results obtained herein confirm their results. Unlike DD-SPH though, our approach directly coupled the two regions of different resolutions without the need for an additional overlap region. Hence, it allowed for a narrower high-resolution region that translates into a reduction in the overall degrees of freedom.

\subsection{A moving colloid in the fluid}\label{subsec:Mov_Cyl}  

One salient attribute of SPH is its ability to model bodies moving freely in the fluid. This brings along the prospect of high-to-low-resolution interfaces that advect with the fluid, i.e, cases in which the high and low resolution regions change dynamically. In the problems studied thus far, the two regions of different resolutions were stationary. Against this backdrop, this section focuses on how the proposed multi-resolution SPH method handles a moving refined region. The vehicle for this study was a 2D ellipsoidal colloid moving in fluid. The high-resolution region moved and its geometry was determined based on the location of the moving colloid. This allowed for an accurately resolved flow in the vicinity of the colloid while saving computational cost via a coarser resolution far from the colloid. 

\begin{figure}[H]
\centering
\includegraphics[scale=0.5]{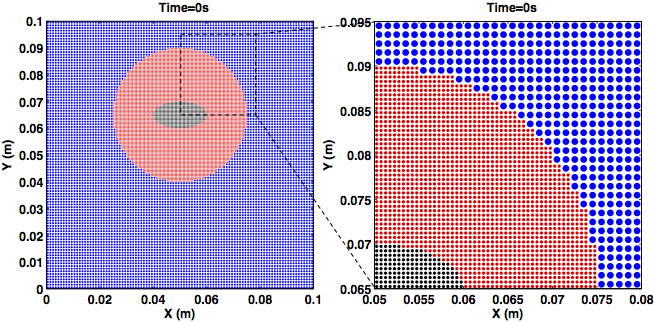}
\caption{Initial configuration of particles with two resolutions for simulating a moving ellipsoid in fluid using the multi-resolution SPH.}
\label{fig:Case4-Ini-Dis-OurSPH-Elli}
\end{figure}

The computational domain was a square box of length $0.1~m$. The length of the ellipsoid's minor axis was $a=0.005~m$; the major axis was $2a$. Initially, the center of the ellipsoid was located at $x=0.05~m$ and $y=0.065~m$. The high-resolution region, for which $\Delta x_H=0.5~mm$, was determined as a moving circular region of radius $0.025~m$ with its center coincident with that of the moving colloid. The low-resolution region with $\Delta x_L=1~mm$ was the remaining fluid domain. The initial setup of the two regions is shown in Figure \ref{fig:Case4-Ini-Dis-OurSPH-Elli}. At $t=0~s$, the ellipsoid was kinematically constrained to move downward as a rigid body with a constant acceleration of $10^{-5}~m/s^{2}$ along the $y$ coordinate. As the ellipsoid reached its maximal translational velocity of $10^{-4}~m/s$, it moved at this constant velocity. Also at $t=0~s$, an additional motion was imposed on the ellipsoid -- it was rotated as a rigid body counter-clockwise at a constant angular acceleration of $1.25\pi\times10^{-4} /s^2$. Once the angular velocity reached $2.5\pi\times10^{-3}/s$, the ellipsoid proceeded to rotate at this constant angular velocity. The no-slip boundary condition was imposed at the boundary of the ellipsoid; periodic boundary conditions were applied at the remaining boundaries. 

\begin{figure}
\centering
\subfigure[Single resolution.]{
\includegraphics[scale=0.65]{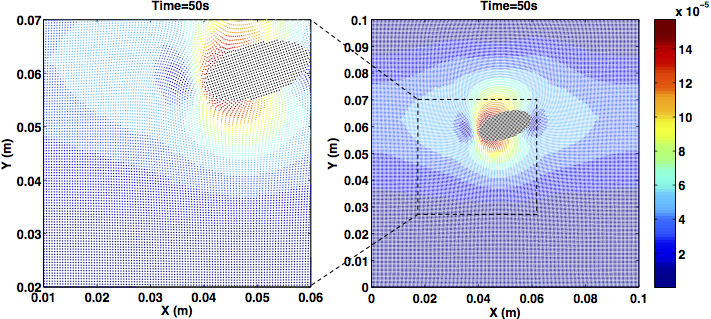}}
\subfigure[Dual resolutions.]{
\includegraphics[scale=0.65]{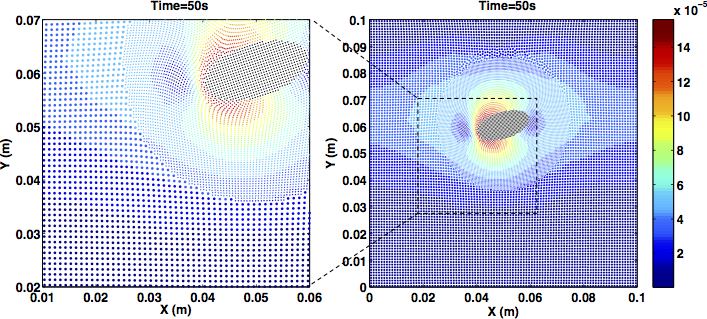}}
\caption{Particle distribution with color correlated to velocity magnitude; snapshot at $t=50~s$ (unit in the color bar: $m/s$).}
\label{fig:Case4-Final-Dis-OurSPH-Elli-50s}
\end{figure}

\begin{figure}
\centering
\subfigure[Single resolution.]{
\includegraphics[scale=0.65]{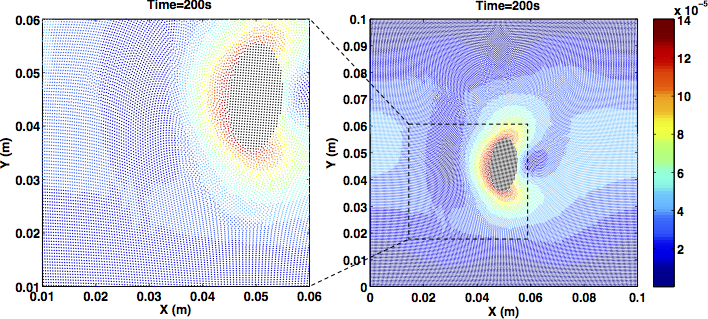}}
\subfigure[Dual resolutions.]{
\includegraphics[scale=0.65]{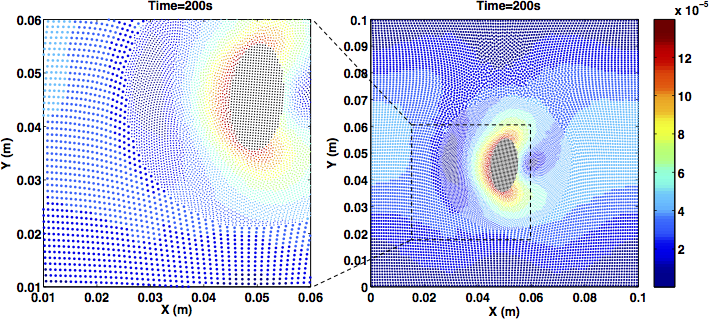}}
\caption{Particle distribution with color correlated to velocity magnitude; snapshot at $t=200~s$ (unit in the color bar: $m/s$).}
\label{fig:Case4-Final-Dis-OurSPH-Elli-200s}
\end{figure}

\begin{figure}
\centering
\subfigure[Single resolution.]{
\includegraphics[scale=0.65]{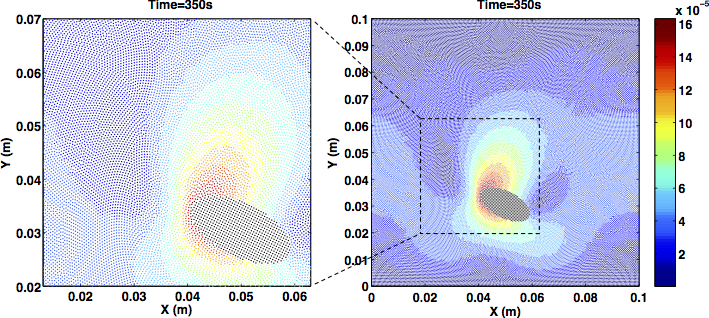}}
\subfigure[Dual resolutions.]{
\includegraphics[scale=0.65]{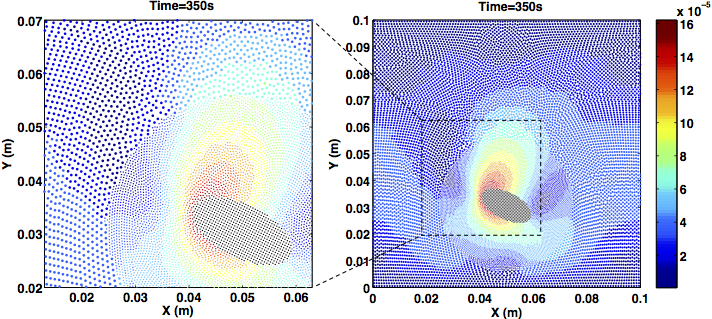}}
\caption{Particle distribution with color correlated to velocity magnitude; snapshot at $t=350~s$ (unit in the color bar: $m/s$).}
\label{fig:Case4-Final-Dis-OurSPH-Elli-350s}
\end{figure}

The dual-resolution solution is compared in terms of accuracy with the single high-resolution counterpart. Figures \ref{fig:Case4-Final-Dis-OurSPH-Elli-50s}--\ref{fig:Case4-Final-Dis-OurSPH-Elli-350s} present at three different times the distribution of SPH particles; their colors are correlated with their velocity magnitude. The quantitative agreement with the single high-resolution SPH insofar the velocity field is concerned indicates that the solution of multi-resolution SPH with a moving refined region is accurate. In addition, there were 16,727 particles used in the multi-resolution SPH, only about 2/5 of 41,600 particles used in the single resolution. 
For each 10-$s$ (physical time) simulation, the computational time of single-resolution solution is 605 $s$, but only 295 $s$ for the multi-resolution solution.

\newpage
\section{Conclusion}
\label{sec:conclusion}

We have presented a new consistent multi-resolution SPH method. It was applied in the context of four different tests: transient Poiseuille flow, flow around a periodic array of cylinders, flow around a rotating cylinder near a moving wall, and an ellipsoidal colloid that translates and rotates in the fluid. The numerical results were compared to analytical, FEM, and consistent single-resolution SPH solutions. We reported good accuracy, second-order convergence, and notable reductions in SPH particle counts. The study was conducted in 2D. Yet, the approach is directly applicable to the 3D case, which insofar efficiency gains are concerned, stands to benefit even more from a multi-resolution approach.

The cornerstone of the proposed method is the second-order consistent SPH discretization employed to ensure the accuracy of the numerical solution when regions of different resolutions are {\textit{directly}} coupled. We found that the accuracy and order of convergence were affected by the ratio of the two subdomains' resolutions ($\Upsilon_{LH}$). The second-order convergence can be maintained for $\Upsilon_{LH} < 2$. For $\Upsilon_{LH} \geq 2$, our convergence study suggests a gradual decay in the convergence order. As an optimal cost--accuracy trade-off, we chose $\Upsilon_{LH}=2$ for all remaining simulations.

A main component of the proposed technique is its splitting and merging strategy for SPH particles leaving or entering the high-resolution region. Both for splitting and merging, the process calls for the state variables to be corrected via the second-order interpolation thus maintaining the second-order accuracy of the consistent SPH discretization. In this setup, the predefined multi-resolution configuration was maintained throughout the simulation despite the SPH particles advecting in and out of these high/low-resolution regions. 

Another noteworthy step of the method is a particle-shifting technique employed to enforce particle regularity. The latter is essential in Lagrangian particle-based methods to ensure numerical stability and accuracy. The shifting vector was computed by accounting for the anisotropic particle distribution at the high-to-low-resolution interface. We slightly shifted particles away from streamlines with hydrodynamic variables corrected to account for their new positions using the second-order interpolation. This strategy avoided a particle-spacing distortion, which in turn kept error propagation in check and enhanced stability without compromising the computational efficiency and Lagrangian nature of the solution.

Finally, two additional points, both pertaining to boundary conditions, were important in the overall accuracy of the approach. First, we imposed no-slip boundary conditions via ghost solid particles. Their velocities were linearly extrapolated from the fluid velocity. The second-order accuracy of this approach matches that of the consistent SPH. Second, a ``smoothed approximation'' was used for calculating SPH particle distance to the boundary, an ingredient that comes into play when imposing no-slip conditions on complex boundaries.

Moving on to aspects related to the computational cost of the proposed method, we first noted that the time step in the explicit integration was determined by $\Delta x_H$ and the handling of the NS viscous term. The proposed multi-resolution method does not introduce any additional constraints for the time step. Second, as different resolutions lead to different kernel lengths, big particles have a larger compact support and hence interact with more neighbor particles near the low-to-high-resolution interface. It leads to the computation of more pair interactions. However, this computation accounts for less than $2\%$ of the computational time in all of our simulations. Third, splitting and merging of SPH particles introduce an unavoidable computational overhead. Upon profiling all our tests, we found that this overhead is less than $1\%$ of the total computational time. This suggests that additional tasks mandatory in a multi-resolution solution lead to an insignificant burden. By comparison, the reduction in number of SPH particles was significant in the 2D setup considered here, with the 3D scenario standing to witness even more dramatic reductions in particle counts.

Looking ahead, a 3D implementation represents a future incremental step that will likely pose limited challenges. The less obvious aspects that remain to be addressed is how to develop consistent second-order methods that vary the resolution continuously, and how to account for the presence of features such as a wall, cables, shells, etc., where a non-spherical kernel might yield an improvement in accuracy and/or efficiency gains.

\section*{Acknowledgments}
This work was supported in part by the 111 China Project (B16003) of and the National Science Foundation of China grants 11290151 and 11221202. The research was made possible in part by the US National Science Foundation grant GOALI-CMMI 1362583.

\section*{References}

\bibliographystyle{elsarticle-num}

\begin{appendix}
\setcounter{secnumdepth}{0}
\section{Appendix: Correction Matrices $\textbf{G}$ and $\textbf{L}$}
\label{sec:Appendix}

The symmetric gradient correction matrix $\textbf{G}_i$ of particle $i$ can be expressed in 2D as:
\begin{equation}\label{equ:gt_exp}
  \textbf{G}_{i}     =   \left(
                          \begin{array}{ccc}
                            G^{11}_i & G^{12}_i \\
                            G^{12}_i & G^{22}_i \\
                          \end{array}
                         \right).
\end{equation}
According to Eq.~\eqref{equ:gt_inv}, the inverse matrix of $\textbf{G}_i$ can be determined by:
\begin{equation}\label{equ:gt_inv_exp}
  \textbf{G}_{i}^{-1} = - \left(
                          \begin{array}{ccc}
                            \sum\limits_j r_{ij}^1\nabla_{i,1}W_{ij}V_{j} & \sum\limits_j r_{ij}^1\nabla_{i,2}W_{ij}V_{j} \\
                            \sum\limits_j r_{ij}^2\nabla_{i,1}W_{ij}V_{j} & \sum\limits_j r_{ij}^2\nabla_{i,2}W_{ij}V_{j} \\
                          \end{array}
                          \right).
\end{equation}
Thus, each component of $\textbf{G}_i$ is computed by inverting the above matrix.

The symmetric Laplacian correction matrix $\textbf{L}_{i}$ of particle $i$ can be expressed in 2D as:
\begin{equation}\label{equ:lt_exp}
  \textbf{L}_{i}     =   \left(
                          \begin{array}{ccc}
                            L_i^{11} & L_i^{12} \\
                            L_i^{12} & L_i^{22} \\
                          \end{array}
                        \right).
\end{equation}
Rewrite Eq.~\eqref{equ:delta_mn} as:
\begin{equation}\label{equ:bt_lt}
\textbf{B}_i
               \left(
                 \begin{array}{c}
                   L_i^{11} \\
                   L_i^{12} \\
                   L_i^{22} \\
                 \end{array}
               \right) \\
 =-\left(
    \begin{array}{c}
      1 \\
      0 \\
      1\\
    \end{array}
  \right),
\end{equation}
where the matrix $\textbf{B}_i$ is expressed as:
\begin{equation}\label{equ:bt_exp}
  \textbf{B}_{i} =  \left(
                          \begin{array}{ccc}
                            B_{i}^{11} & B_{i}^{12} & B_{i}^{13} \\
                            B_{i}^{21} & B_{i}^{22} & B_{i}^{23} \\
                            B_{i}^{31} & B_{i}^{32} & B_{i}^{33} \\
                          \end{array}
                        \right).
\end{equation}
The components of this $3\times3$ matrix $\textbf{B}_i$ are defined as:
\begin{equation}
\begin{split}
& B_{i}^{11} = \sum\limits_j (A_{i}^{111}e_{ij}^1+A_{i}^{211}e_{ij}^2+r_{ij}^1 e_{ij}^1)(e_{ij}^1\nabla_{i,1}W_{ij})V_j\\
& B_{i}^{12} = \sum\limits_j (A_{i}^{111}e_{ij}^1+A_{i}^{211}e_{ij}^2+r_{ij}^1 e_{ij}^1)
               (e_{ij}^1\nabla_{i,2}W_{ij}+e_{ij}^2\nabla_{i,1}W_{ij})V_j\\
& B_{i}^{13} = \sum\limits_j (A_{i}^{111}e_{ij}^1+A_{i}^{211}e_{ij}^2+r_{ij}^1 e_{ij}^1)(e_{ij}^2\nabla_{i,2}W_{ij})V_j\\
& B_{i}^{21} = \sum\limits_j (A_{i}^{112}e_{ij}^1+A_{i}^{212}e_{ij}^2+r_{ij}^1 e_{ij}^2)(e_{ij}^1\nabla_{i,1}W_{ij})V_j\\
& B_{i}^{22} = \sum\limits_j (A_{i}^{112}e_{ij}^1+A_{i}^{212}e_{ij}^2+r_{ij}^1 e_{ij}^2)
               (e_{ij}^1\nabla_{i,2}W_{ij}+e_{ij}^2\nabla_{i,1}W_{ij})V_j\\
& B_{i}^{23} = \sum\limits_j (A_{i}^{112}e_{ij}^1+A_{i}^{212}e_{ij}^2+r_{ij}^1 e_{ij}^2)(e_{ij}^2\nabla_{i,2}W_{ij})V_j\\
& B_{i}^{31} = \sum\limits_j (A_{i}^{122}e_{ij}^1+A_{i}^{222}e_{ij}^2+r_{ij}^2 e_{ij}^2)(e_{ij}^1\nabla_{i,1}W_{ij})V_j\\
& B_{i}^{32} = \sum\limits_j (A_{i}^{122}e_{ij}^1+A_{i}^{222}e_{ij}^2+r_{ij}^2 e_{ij}^2)
               (e_{ij}^1\nabla_{i,2}W_{ij}+e_{ij}^2\nabla_{i,1}W_{ij})V_j\\
& B_{i}^{33} = \sum\limits_j (A_{i}^{122}e_{ij}^1+A_{i}^{222}e_{ij}^2+r_{ij}^2 e_{ij}^2)(e_{ij}^2\nabla_{i,2}W_{ij})V_j\\
\end{split}
\end{equation}
Here, the components of the symmetric third-order tensor $\textbf{A}_i$ are determined by:
\begin{equation}
\begin{split}
  A_{i}^{111} = \sum\limits_j r_{ij}^1 r_{ij}^1 (G_{i}^{11}\nabla_{i,1}W_{ij}+G_{i}^{12}\nabla_{i,2}W_{ij})V_j\\
  A_{i}^{112} = \sum\limits_j r_{ij}^1 r_{ij}^2 (G_{i}^{11}\nabla_{i,1}W_{ij}+G_{i}^{12}\nabla_{i,2}W_{ij})V_j\\
  A_{i}^{122} = \sum\limits_j r_{ij}^2 r_{ij}^2 (G_{i}^{11}\nabla_{i,1}W_{ij}+G_{i}^{12}\nabla_{i,2}W_{ij})V_j\\
  A_{i}^{211} = \sum\limits_j r_{ij}^1 r_{ij}^1 (G_{i}^{12}\nabla_{i,1}W_{ij}+G_{i}^{22}\nabla_{i,2}W_{ij})V_j\\
  A_{i}^{212} = \sum\limits_j r_{ij}^1 r_{ij}^2 (G_{i}^{12}\nabla_{i,1}W_{ij}+G_{i}^{22}\nabla_{i,2}W_{ij})V_j\\
  A_{i}^{222} = \sum\limits_j r_{ij}^2 r_{ij}^2 (G_{i}^{12}\nabla_{i,1}W_{ij}+G_{i}^{22}\nabla_{i,2}W_{ij})V_j\\
\end{split}
\end{equation}
Given $\textbf{A}_i$ and $\textbf{B}_i$, $L_i^{11}$, $L_i^{12}$, and $L_i^{22}$ can be obtained by solving Eq.~\eqref{equ:bt_lt}.

Note that computing $\textbf{G}_i$ and $\textbf{L}_{i}$ for each particle $i$ requires no additional neighbor information than the standard SPH operators.

\end{appendix}

\end{document}